\documentclass[a4,12pt,twoside]{article}
\usepackage[]{graphicx}\usepackage[]{color}
\makeatletter
\def\maxwidth{ %
  \ifdim\Gin@nat@width>\linewidth
    \linewidth
  \else
    \Gin@nat@width
  \fi
}
\makeatother

\definecolor{fgcolor}{rgb}{0.345, 0.345, 0.345}

\usepackage{framed}
\makeatletter
 {\par\unskip\endMakeFramed%
 \at@end@of@kframe}
\makeatother

\definecolor{shadecolor}{rgb}{.97, .97, .97}
\definecolor{messagecolor}{rgb}{0, 0, 0}
\definecolor{warningcolor}{rgb}{1, 0, 1}
\definecolor{errorcolor}{rgb}{1, 0, 0}
\newenvironment{knitrout}{}{} 

\usepackage{alltt}

\usepackage[lmargin=1.25in,rmargin=1.25in,tmargin=1.25in,bmargin=1.25in]{geometry}
\usepackage{fancyhdr} 
\usepackage{times}
\usepackage{mathptmx} 
\usepackage[compact]{titlesec}
\usepackage[page]{appendix}
\usepackage[utf8]{inputenc}										
\usepackage{booktabs,multirow,rotating,pdflscape,array,numname,dcolumn}	 											
\usepackage{graphics,epsfig,float,amsmath,amsfonts,longtable,graphicx,endnotes}		
\usepackage{natbib}                                                 
\setcitestyle{aysep={}}
\usepackage[table]{xcolor}
\usepackage[hidelinks,bookmarksnumbered]{hyperref}
\usepackage[nomain,acronym,xindy,toc,nonumberlist]{glossaries}
\usepackage[xindy]{imakeidx} 
\usepackage{microtype}
\usepackage{authblk} 
\usepackage{xparse}
\DeclareDocumentCommand{\newdualentry}{ O{} O{} m m m m } {
  \newglossaryentry{gls-#3}{name={#5},text={#5\glsadd{#3}},
    description={#6},#1
  }
  \makeglossaries
  \newacronym[see={[Glossary:]{gls-#3}},#2]{#3}{#4}{#5\glsadd{gls-#3}}
}

\newcolumntype{L}[1]{>{\raggedright\arraybackslash}p{#1}}
\newcolumntype{C}[1]{>{\centering\arraybackslash}p{#1}}

\newcommand{\Ev}[1]{\operatorname{E}\left[ #1 \right]}

\makeatletter
\newcommand{\distras}[1]{\mathbin{\overset{#1}{\kern\z@\sim}}}%
\newsavebox{\mybox}\newsavebox{\mysim}
\newcommand{\distas}[1]{%
  \savebox{\mybox}{\hbox{\kern3pt$\scriptstyle#1$\kern3pt}}%
  \savebox{\mysim}{\hbox{$\sim$}}%
  \mathbin{\overset{#1}{\kern\z@\resizebox{\wd\mybox}{\ht\mysim}{$\sim$}}}%
}
\makeatother

\definecolor{green}{HTML}{282828}
\definecolor{red}{HTML}{808080}
\definecolor{blue}{HTML}{555555}
\definecolor{purple}{HTML}{858585}

\let\footnote=\endnote

\makeglossaries

\titlespacing{\section}{0pt}{1ex}{1ex}
\titlespacing{\subsection}{0pt}{0.5ex}{0ex}
\titlespacing{\subsubsection}{0pt}{0.2ex}{0ex}

\title{Decomposing ethnic achievement gaps across multiple levels of analysis and for multiple ethnic groups}
\date{January 2022}
\author[1]{Beatriz Gallo Cordoba}
\author[2]{George Leckie}
\author[2]{William J.  Browne}
\affil[1]{Centre for Youth Policy and Education Practice,  Faculty of Education,  Monash University}
\affil[2]{Centre for Multilevel Modelling,  School of Education, University of Bristol}

\newdualentry{moe} 
  {MoE}            
  {Ministry of Education}  
  {Colombian national-level organisation that is responsible for education} 

\newdualentry{icfes} 
  {Icfes}            
  {Colombian National Assessment Institute}  
  {The national-level institution that is in charge of educational assessment in Colombia, including the administration of SABER 11 exam} 

\newdualentry{bc} 
  {BC}            
  {Buscando Colegio}  
  {It literally translates ``searching for school''. This is the database that contains venue-level information about all colombian schools} 

\newdualentry{sisben} 
  {Sisben}            
  {System for the Selection of Social Programs' Beneficiaries}  
  {An instrument for selecting public policy beneficiaries in Colombia} 

\newdualentry{natplan} 
  {NDP}            
  {National Development Plan}  
  {A four-year plan outlining the central government programs and investments in order to reach development goals} 

\newglossaryentry{saber11} 
{
  name=  SABER 11,
  description={Exam that all Colombian students must present at the end of compulsory education}
}

\newacronym{dane}
	{DANE} 
	{National Bureau of Statistics} 
	
\newacronym{sd}
	{SD} 
	{standard deviations} 
	
\newacronym{vcm}
	{VCM} 
	{variance component model} 
	
\newacronym{icc}
	{ICC} 
	{intraclass correlation} 

\newacronym{ses}
	{SES} 
	{socio-economic status} 

\newacronym{pca}
	{PCA} 
	{principal component analysis} 

\newacronym{ca}
	{CA} 
	{correspondence analysis} 

\newacronym{mca}
	{MCA} 
	{multiple correspondence analysis} 

\newacronym{fa}
	{FA} 
	{factor analysis} 

\newacronym{ir}
	{IR} 
	{item response} 
	
\newacronym{uv}
	{UV} 
	{underlying variable} 

\newacronym{lta}
	{LTA} 
	{latent trait analysis}  

\newacronym{lca}
	{LCA} 
	{latent class analysis}  
		
\newacronym{dnp}
	{DNP} 
	{Colombian Department for National Planning}  

\newacronym{oecd}
	{OECD} 
	{Organisation for Economic Co-operation and Development}  

\newacronym{ols}
	{OLS} 
	{Ordinary Least Squares}  

\newacronym{esrc}
	{ESRC} 
	{Economic and Social Research Council}  

\newacronym{sem}
	{SEM} 
	{structural equation modelling}  

\newacronym{pisa}
	{PISA} 
	{Programme for International Student Assessment}  

\newacronym{naep}
	{NAEP} 
	{National Assessment of Educational Progress}  

\newacronym{timss}
	{TIMSS} 
	{Trends in International Mathematics and Science Study}  

\newacronym{pirls}
	{PIRLS} 
	{Progress in International Reading Literacy Study}  

\newacronym{iea}
	{IEA} 
	{International Association for the Evaluation of Educational Achievement}  

\newacronym{nces}
	{NCES} 
	{National Centre for Education Statistics in the USA}  

\newacronym{fsm}
	{FSM} 
	{Free School Meal}  

\newacronym{aic}
	{AIC} 
	{Akaike information criterion}  

\newacronym{bic}
	{BIC} 
	{Bayesian information criterion}  

\newacronym[]{la}
	{district} 
	{school district}  

\newacronym{eric}
	{ERIC} 
	{Education Resources Information Center}  

\IfFileExists{upquote.sty}{\usepackage{upquote}}{}
\begin{document}

\maketitle

\noindent This is the accepted version of the following journal article:
  
\noindent Gallo Cordoba B, Leckie G, Browne W.J. Decomposing Ethnic Achievement Gaps across Multiple Levels of Analysis and for Multiple Ethnic Groups. \textit{Sociological Methodology}.  pp. 1-31.  Copyright\copyright \ [June 2022] (American Sociological Association).  doi: \href{https://doi.org/10.1177/00811750221099503}{10.1177/00811750221099503}

\pagebreak




\section*{Acknowledgement of funding}
\noindent This work was supported by the Economic and Social Research Council [grant numbers ES/J50015X/1; ES/R010285/1].

\section*{Address for correspondece}
\noindent Dr. Beatriz Gallo Cordoba

\begin{description}
\item[Address:] Faculty of Education\\
Level 1, Learning and Teaching Building.\\Monash University, Clayton campus.\\
19 Ancora Imparo Way\\
Clayton VIC 3800\\
Australia
\item[T:] +61 3 9905 2768
\item[E:] \href{mailto:beatriz.gallocordoba@monash.edu}{beatriz.gallocordoba@monash.edu}
\end{description}
\vfill
\pagebreak

\begin{center}
\textbf{\Large{Decomposing ethnic achievement gaps across multiple levels of analysis and for multiple ethnic groups}}
\end{center}

\section*{Abstract}

Ethnic achievement gaps are often explained in terms of student and school factors. The decomposition of these gaps into their within- and between-school components has therefore been applied as a strategy to quantify the overall influence of each set of factors. Three competing approaches have previously been proposed, but each is limited to the study of student-school decompositions of the gap between two ethnic groups (e.g., White and Black). We show that these approaches can be reformulated as mediation models facilitating new extensions to allow for additional levels in the school system (e.g., classrooms, school districts, or geographic areas) and multiple ethnic groups (e.g., White, Black, Hispanic, and Asian). We illustrate these extensions using administrative data for high-school Colombian students and we highlight the increased substantive insights and nuanced policy implications that they afford.

\subsection*{Keywords}
\noindent Achievement gaps, mediation analysis, effect decomposition, multiple-level decomposition, multi-group decomposition.

\section{Introduction}\label{sx:intro}

Ethnic achievement gaps, usually defined as the difference between the mean test scores of two groups of students (e.g. Black and White) have been widely explored in education, psychology, sociology and other behavioural sciences. This literature suggests that these gaps are the result of differences in a wide range of variables defined at the student, school and higher levels of the education system \citep{Coleman1966,Rothstein2004C,Mohammadpour2014,Bidwell1975S,Wenglinsky2009H}. One way of addressing the challenge of finding the variables that potentially drive achievement gaps is to know where to start looking for them.

A natural approach is therefore to decompose the overall achievement gap into its separate component parts operating at each level of the education system. Once this has been achieved, variables at the corresponding levels (e.g.  student, school and school-district characteristics) can be considered as potential explanatory variables of the achievement gap. For example, if most of the achievement gap is due to differences within schools, the search might start with student-level variables; while if most of the achievement gap is a consequence of differences between schools, school-level variables might be prioritised, including both schools' characteristics and composition. It is also of interest how this within- and between-school decomposition changes over cohorts and school grades, and to compare  across regions and countries. Identifying the contribution of different levels of the education system to the achievement gap can also help to prioritise policy efforts at specific levels and to establish what level of the education system may be held accountable for disparities in achievement.

Within the context of studying the Black-White achievement gap in the US, three competing methodological approaches have been proposed.  We refer to these as Approach 1 \citep{Cook2000,Fryer2004a,Fryer2006a}, Approach 2  \citep{Hanushek2007} and Approach 3  \citep{Page2008,Reardon2008a}. Authors applying Approaches 1 and 3  argued that between-school differences explain at most 40\% of the White-Black achievement gap in the US \citep{Page2008,Cook2000,Fryer2004a,Fryer2006a,Reardon2008a}, while authors applying Approach 2 argued that these differences explain 70\% of the achievement gap \citep{Hanushek2007}.

These current approaches are limited in at least two ways. First, they do not consider the role of other levels of the school system, such as \glspl{la}. Second, they are restricted to a binary comparison, ignoring the role of other minority groups. These omissions to features of real data may hinder the pertinence and relevance of such a gap decomposition analysis. For example, using the existing approaches can result in directing the attention to school-level policies such as providing teacher training, while a more relevant policy may focus on \gls{la}-level interventions such as reconsidering the way resources are allocated. 

The main contribution of this paper is to extend the three current decomposition approaches to consider additional levels of the school system (e.g. \glspl{la}) and multiple ethnic groups (e.g. White, Black, Asian, Hispanic) using mediation analysis as a framework to facilitate such extensions. 

We first show that the current achievement gap decomposition approaches are mathematically equivalent to a mediation problem. We then take advantage of this equivalence to extend the current decomposition approaches. The reader should note that we do not use mediation analysis in the traditional sense \citep{Vanderweele2009,Baron1986}, but merely as a device to facilitate the effect decomposition.  \citet{Hou2014A} proposed using mediation analysis as a general framework for effect decomposition. However, the use of mediation analysis as a methodological tool to extend the ethnic achievement gap decomposition approaches has not been previously considered in the literature.   Although this is not the only option to derive the extensions presented in this paper (e.g.  deriving the mathematical formulation directly is another option), it facilitates the presentation of these extensions.  We apply these extensions to the three approaches to study ethnic achievement gaps in Colombia to illustrate the extensions and their importance.

The following section shows how the three existing ethnic gap decomposition approaches are mathematically equivalent to a mediation problem.  Section \ref{sx:agdata} describes the data and presents the results of applying the existing two-ethnic-group and two-level decomposition approaches. Sections \ref{sx:multlevels} and \ref{sx:multgroups} extend the three current methodological approaches to consider additional levels of the school system (e.g. \glspl{la}) and multiple ethnic groups, respectively. These sections also include an illustration of such approaches using the data described in section \ref{sx:agdata}.  Section \ref{sx:serialmed} explores an alternative mediation approach for these extensions.  Section \ref{sx:agdisc} discusses and concludes.

\section{Approaches for the Within and Between-School Gap Decomposition}\label{sx:aglitreview}

 \citet{Reardon2008a} and \citet{Page2008} have previously reviewed the three existing methodological approaches for the ethnic achievement gap decomposition and discussed their interpretation. This section summarizes these approaches from a mediation analysis perspective, in preparation for the extensions  in sections \ref{sx:multlevels} and \ref{sx:multgroups}, which present our main contribution.  Here,  we understand mediation analysis merely as a tool that allows us to decompose the effect of ethnicity (i.e.  the ethnic achievement gap) into different components.  Nonetheless,  using the mediation analysis framework is not essential but a means to deriving the extensions we propose. 

Consider the single-level linear regression model 
\begin{equation}\label{eq:gap}
\begin{matrix}
y_{ij}  = \alpha +  \beta M_{ij}+{e_y}_{ij}\\
\\
{e_y}_{ij} \sim  N\left(0,\Omega\right)\\
\end{matrix}
\end{equation}

\noindent where the dependent variable $y_{ij}$ is the  test score of student $i$ attending school $j$ and the independent variable $M_{ij}$ is a dummy variable that equals one for ethnic minority students and 0 otherwise. In this case, $\beta$ represents the average difference in  test scores between minority and White students; the overall ethnic achievement gap.

The difference in the average test scores between White and minority students within and between schools can be estimated using the hybrid-effect model\footnote{For more on this type of model and the interpretation of the estimated parameters please refer to \citet{Castellano2014,Mundlak1978,Brincks2017}.} 
\begin{equation}\label{eq:scmino}
\begin{array}{rcl}
y_{ij}&=&\alpha+\beta^W\left(M_{ij}-\overline{M}_{.j}\right) + \beta^{S}\overline{M}_{.j}+{e_y}_{ij} \\
 &=& \alpha + \beta^W M_{ij} + \left(\beta^{S}-\beta^W\right) \overline{M}_{.j}+{e_y}_{ij} \\
  &=& \alpha + \beta^W M_{ij} + \beta^{C}\overline{M}_{.j}+{e_y}_{ij} \\
\end{array}
\end{equation}

\noindent where $\overline{M}_{.j}$ is the proportion of minority students in school $j$. Thus, $\beta^W$ represents the within-school achievement gap; the average difference in test scores between white and minority students attending the same school. Additionally, $\beta^{S}$ represents the between-school gap; the average difference in test scores between schools that only serve White students  ($\overline{M}_{.j} =0$) and schools that only serve minority students ($\overline{M}_{.j} =1 $). The last line in  \eqref{eq:scmino} illustrates the equivalence between the hybrid-effect model and the contextual effect model. In this last type of model $\beta^{C}=\beta^{S}-\beta^W$ represents the school contextual effect of ethnicity; the effect of attending a school with a larger proportion of minority students, over and above the effect of the students' own ethnicity. 

\begin{figure}[hbt]
\begin{center}
\includegraphics[scale=0.65]{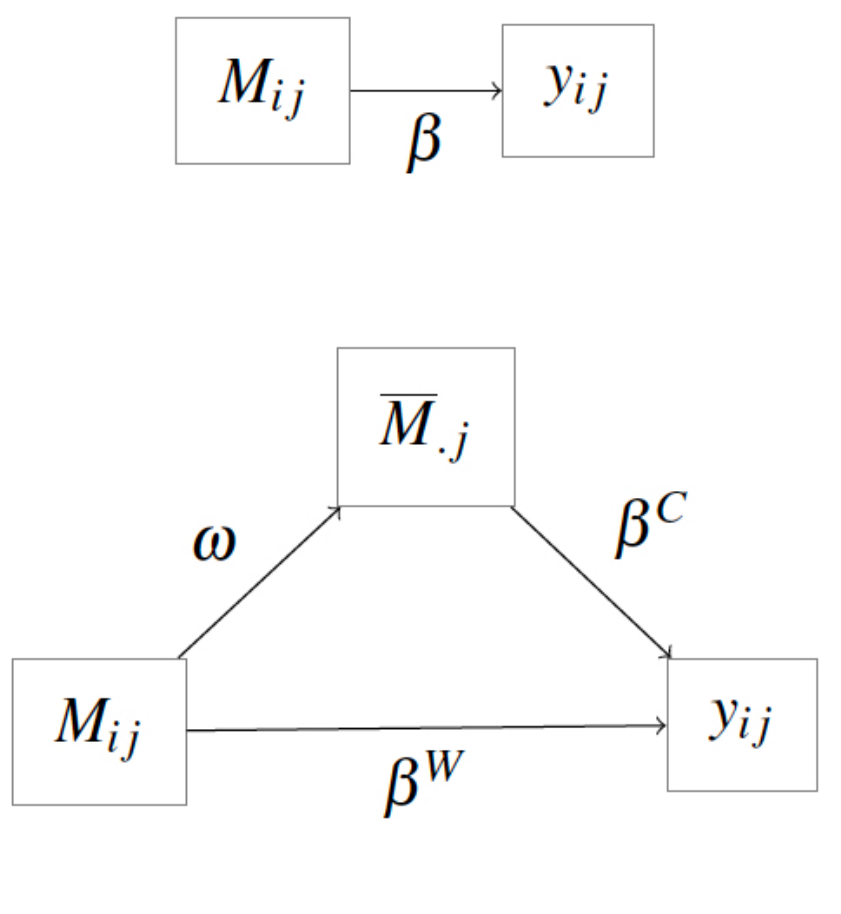}
\caption{Model for the achievement gap decomposition into its within- and between-school components}\label{fg:medmodsim}
\end{center}
\end{figure}

To translate the three  achievement gap decomposition approaches into the mediation analysis framework, it is enough to consider a model in which the ethnic composition of each school $\overline{M}_{.j}$ mediates the relationship between ethnicity $M_{ij}$ and achievement $ y_{ij}$, as we show in Figure \ref{fg:medmodsim}.

The upper part of Figure \ref{fg:medmodsim} describes the total-effect model  \eqref{eq:gap}. The lower part of Figure \ref{fg:medmodsim} represents the outcome and mediation models. The outcome model is the contextual effect model \eqref{eq:scmino}. Besides, the mediation equation is

\begin{equation}\label{eq:mods}
\overline{M}_{.j}=\gamma_S + \omega M_{ij} + {e_S}_{.j}, \forall i \in j
\end{equation}

\noindent where the school proportion of minority students $\overline{M}_{.j}$ is repeated for all students $i$ in school $j$.  Here, $\gamma_S$ represents the average school proportion of minority students experienced by White students, and $\omega$ represents the difference in the average school proportion of minority students experienced by White and minority students. If $\omega~=0$, White and minority students attend schools with the same proportion of minority students, while if $\omega~=1$, White students only attend schools with White students and minority students only attend schools with minority students. As we show in more detail below,  model \eqref{eq:mods} is a simple linear regression model that is used as a device to compute  $\omega$. Given its interpretation, $\omega$ is equivalent to the variance ratio index of segregation, as noted by \citet{Reardon2008a}.

In the mediation analysis context, the total effect $\beta$ is decomposed into a direct effect $\beta^W$ and an indirect effect $\omega\beta^{C} $, such that

\begin{equation}\label{eq:mdec}
\beta= \beta^W + \omega\beta^{C}
\end{equation}

Here, the direct effect is the within-school gap $\beta^W$ and the indirect effect $\omega\beta^{C}$ represents how much more minority students are affected by the school contextual effect of ethnicity  $\beta^{C}$, in comparison to White students, given their average additional exposure to minority students  $\omega$. 

Different strategies can be used to derive confidence intervals for the direct and indirect effect,  as well as for the different components of the gap discussed below,  including bootstrapping and Monte Carlo methods \citep{MacKinnon2004}.  For the application in this paper,  we estimate 95\%, 99\% and 99.9\% Monte Carlo confidence intervals for the proportion of the gap that each of its components represents.  The confidence intervals were implemented via \citet{Selig2008}.  The advantage of Monte Carlo methods is that they only require  the estimated parameters and their standard errors, which can be estimated using cluster-robust estimators to account for the nested nature of the data  \citep{Preacher2012}.   The supplemental material includes R and Stata code to implement the proposed decompositions.

The following subsections review the three decomposition approaches mentioned in section \ref{sx:intro}  and consider how they can fit into a mediation analysis framework.

\subsection{Approach 1: {\citet{Cook2000} and \citet{Fryer2006a,Fryer2004a}}}
\citet{Cook2000} and \citet{Fryer2006a,Fryer2004a} used a version of the Kitagawa-Oaxaca-Blinder decomposition \citep{Blinder1973BW,Oaxaca1973,Kitagawa1955} that included school fixed effects, allowing them to identify achievement gaps within schools, after considering differences in gender and parental education. This model can be written as

\begin{equation}\label{eq:fixdmino}
y_{ij}=\alpha_j+\beta^WM_{ij}+{e_y}_{ij} 
\end{equation}

\noindent where $\alpha_j$ is a dummy variable (fixed effect) for  school $j$. In this case, $\beta^W$ is the within-school ethnic achievement gap; i.e. the difference in mean test scores between White and minority students attending the same school. \citet{Cook2000} and \citet{Fryer2006a,Fryer2004a} attributed $\beta^W$ to within-school differences and the remainder of the gap ({$\beta - \beta^W$}) to between-school differences, where $\beta$ is the overall gap estimated by \eqref{eq:gap}. In the mediation analysis framework, this difference is understood as an alternative to \eqref{eq:mdec} to calculate the indirect effect (of schools, in this case). 

Therefore, the decomposition approach 1 is equivalent to \eqref{eq:mdec}, and  \citet{Cook2000} and \citet{Fryer2006a,Fryer2004a} attributed $\beta^W$ to the within-school component of the gap and $\omega\beta^{C} $ to the between-school component of the gap. The expression in \eqref{eq:mdec} shows that \citet{Cook2000}'s and \citet{Fryer2006a,Fryer2004a}'s interpretation implies that eliminating the within-school gap $\beta^W~=0$ would leave the contribution of the between-school differences to the gap unaffected. 

This statement is true in a scenario without segregation ($\omega~=0$), where the ethnic achievement gap $\beta$ is fully explained by the within-school achievement gap ($\beta=~\beta^W$). In this case, there are no differences in the average test scores of schools typically attended by White and minority students. However, in a scenario with segregation ($\omega~\neq0$), suppressing the within-school ethnic achievement gap ($\beta^W=~0$) leads to changes in ethnic differences  between schools that the between-school component of this decomposition does not capture \citep{Reardon2008a,Hanushek2007}. Therefore, approach 1 decomposes the achievement gap into its direct and indirect effects; the within-school gap $\beta^W$ and the effect of segregation $\omega\beta^{C}$. 

\subsection{Approach 2: {\citet{Hanushek2007}}}
Challenging \citet{Fryer2006a,Fryer2004a}, \citet{Hanushek2007} deduced a mathematical formulation to write the Black-White achievement gap as a weighted sum of differences within- and between-schools. As \citet{Reardon2008a} showed, \citet{Hanushek2007} used the within- and between-school gaps, which can be recovered from the hybrid-effect model \eqref{eq:scmino}.  
\citet{Reardon2008a} also showed that \citet{Hanushek2007}'s formula is equivalent to

\begin{equation}\label{eq:decomp}
\beta=\left(1-\omega\right)\beta^W+\omega\beta^{S}, \omega=\frac{\mathbf{Var}\left(\overline{M}_{.j}\right)}{\mathbf{Var}\left(M_{ij}\right)}
\end{equation} 

Then, \citet{Hanushek2007}  argued that the contribution of the within-school achievement gap $\beta^W$ to the overall gap $\beta$ is $\left(1-\omega\right)\beta^W$ and the contribution of the between-school differences $\beta^{S}$ to the overall achievement gap is $\omega\beta^{S}$, as had already been shown in more general contexts \citep{Cronbach1976,Burstein1980}.

Translating this approach to the mediation analysis framework requires recognising that differences between schools are the sum of differences within schools and the contextual effect of ethnicity. In other words, differences between schools arise not only because of the effect of studying with a larger proportion of minority students ($\beta^{C}$, a school-level effect) but also because of differences between White and minority students across all schools ($\beta^W$). Equivalently, this can be expressed as $\beta^{C}~=\beta^{S}~-\beta^{W}$.

Recognising this in \eqref{eq:mdec} leads to \eqref{eq:decomp}.  Accordingly, the ethnic achievement gap $\beta$ is a weighted\footnote{\citet{Reardon2008a} noticed the equivalence between $\omega$ in \eqref{eq:decomp} and $\omega$ calculated as {$\Ev{\overline{M}_{.j}|M_{ij}=1}-\Ev{\overline{M}_{.j}|M_{ij}=0}$}.} sum of differences within schools $\beta^W$ and between schools $\beta^{S}$. As \citet{Reardon2008a} noticed, in a scenario without segregation ($\omega=0$), this decomposition is equivalent to \citet{Cook2000}'s and \citet{Fryer2006a,Fryer2004a}'s, with $\beta=\beta^W$. If there is segregation ($\omega~\neq0$), suppressing the within school ethnic achievement gap ($\beta^W~=0$) implies changes in both, the within-school $\left(1-\omega\right) \beta^W$ and the between-school $\omega~\beta^{S}$ components of the gap, which would then be 0 and $\omega\beta^{C}$, respectively. 

\subsection{Approach 3: {\citet{Reardon2008a} and \citet{Page2008}}}

Trying to conciliate \citet{Fryer2006a,Fryer2004a} and \citet{Hanushek2007}, \citet{Reardon2008a} argued that the overall achievement gap can be decomposed into three different components: one that can be attributed to  differences within schools, a second one that is due to differences between schools and a third one that is a combination of both and therefore cannot be uniquely assigned to either of those. This decomposition is based on the contextual-effect model  \eqref{eq:scmino}.

This is equivalent to further separating  the contribution of between-school differences to the overall gap, acknowledging that {$\beta^{S}= \beta^{W} + \beta^{C}$} in \eqref{eq:decomp}, which leads to

\begin{equation}\label{eq:reardon}
\beta= \left(1-\omega\right) \beta^W + \omega\beta^{W}+ \omega\beta^{C}
\end{equation}

Under \citet{Reardon2008a}'s interpretation, $\left(1-\omega\right)\beta^W$  is the within-school component of the gap, $\omega\beta^{C}$ is the between-school component, due to segregation, and $\omega\beta^{W}$ is the part of the gap that cannot be unambiguously attributed to either of them.

To interpret the differences between approaches 1 and 2, \citet{Reardon2008a} and \citet{Page2008} proposed analysing the type of policies that would be required to eliminate ethnic differences within and between schools. If eliminating a component of the gap requires a between-school intervention, it can be attributed to the between-school gap. Conversely, if it requires a within-school intervention, it can be attributed to the within-school component of the gap. Their argument is based on three different policies: 

\begin{enumerate}
\item Eliminating the within-school gap ($\beta^W~=0$) while leaving segregation ($\omega$) and the school mean achievement unchanged\footnote{Please note that in this scenario, leaving the school mean unchanged requires an increase of $\beta^W$ in the school contextual effect of ethnicity $\beta^{C}$.}: This within-school effort would lead to a decrease of {$\left(1-\omega\right)\beta^W$} in the overall gap $\beta$, which now is {$\beta=\omega\beta^{C}$}, composed only of differences between schools. Therefore, \citet{Reardon2008a} and \citet{Page2008} argued that {$\left(1-\omega\right)\beta^W$} is the part of the gap that can `unambiguously' be attributed to within-school differences, in agreement with \citet{Hanushek2007}.

\item Eliminating segregation ($\omega~=0$): This between-school effort would lead to a decrease of {$\omega\beta^{C}$} in the overall gap, which would now only be the within-school gap $\beta=~\beta^W$, as argued by \citet{Fryer2006a,Fryer2004a}. Therefore, \citet{Reardon2008a} and \citet{Page2008} interpreted {$\omega\beta^{C}$} as the part of the gap that is attributable to segregation and `unambiguously' to between-school differences.

\item Eliminating the relationship between the school ethnic composition and its mean achievement {($\beta^W+\beta^{C}=0$)} while keeping the within-school gap $\beta^W$ and segregation $\omega$ at the same levels\footnote{Please note that such a  policy is only possible if {$\beta^{W}=-\beta^{C}$}; i.e. if the within-school gap offsets the school contextual effect of ethnicity.}: This between-school effort would lead to a reduction of {$\omega\left(\beta^W+\beta^{C}\right)$} in the overall gap $\beta$, which would now be  {$\beta=\left(1-\omega\right)\beta^W$}; the `unambiguously' within-school component of the gap, which \citet{Reardon2008a} and \citet{Page2008} argued is in agreement with \citet{Hanushek2007}'s decomposition. 
\end{enumerate}

Using these policy scenarios, \citet{Reardon2008a} and \citet{Page2008} argued that $\omega\beta^W$ cannot be `unambiguously attributed' to either the within- or the between-school component of the gap because it is an interaction between within-school differences $\beta^W$ and segregation $\omega$, which can be changed through interventions either within or between schools. The difference between approaches 1 and 2,  they argued, is that approach 1 attributes $\omega\beta^W$  to the within-school component, while approach 2 attributes it to the between-school component. This argument can be questioned\footnote{This proposition does not question the decomposition itself, but the argument used to decide how to interpret each of the components.} by noticing that the `unambiguously within-school' component of the gap {$\left(1-\omega\right)\beta^W$} could also be eliminated through a between-school intervention: completely segregating schools ({$\omega=1$}). We argue that what \citet{Reardon2008a} and \citet{Page2008} call the `ambiguous' component of the gap  ($\omega\beta^{W}$) is the contribution of segregation through differences in the schools' student intake and $\omega\beta^{C}$ is the contribution of segregation through the school contextual effect of ethnicity. Adding these two results in the contribution of between-school differences to the overall ethnic achievement gap, as \citet{Hanushek2007} argued. 

Additionally, \citet{Page2008} combined \citet{Reardon2008a}'s approach with the Kitagawa-Oaxaca-Blinder decomposition to incorporate covariates in the model, including indicator variables for other ethnicities. They are the first authors to consider other ethnicities, however, their approach did not acknowledge that the influence of other groups can be part of the between-school component of the Black-White achievement gap, as discussed in section \ref{sx:multgroups}.  We will now turn our attention to the application of interest in this paper.

\section{Data and Initial Application}\label{sx:agdata}
At the end of secondary education, all Colombian students take a standardised exam, called \gls{saber11}. The compulsory nature of the exam implies that the data resulting from its administration is effectively a census of 11th grade (age 16/17) students in Colombia, including those attending private and public (state-funded) schools. The nation-wide educational authority in Colombia is the Ministry of Education, but there are smaller administrative divisions --\glspl{la} or {\it{entidades territoriales certificadas}}-- that are responsible for the management of education within their geographical boundaries.

The main ethnic groups in Colombia are: the white-mestizo group  (European-descendants who have mixed with other ethnic groups, 85.9\% of the population), Afrocolombians (10.6\%) and Indigenous (3.4\%), while other ethnic minorities account for 0.01\% of the population \citep{DANE2007}. Afrocolombian, Indigenous and other minority students are grouped together for the applications presented here and in section \ref{sx:appml}. Hence, the overall ethnic achievement gap is defined as the difference in the average test scores between  White and minority students.

We focus on the \gls{saber11} math test scores for 2011, which included 458,947 students,  8,039 schools and 94 \glspl{la}. Of all students, 6.7\% self-recognise as belonging to an ethnic minority. Only 34.9\% of the schools serve both white and minority students, and there is large variation in the proportion of minority students within schools, regardless of their size.  The math test scores have been normalised to have a mean of zero and a standard deviation of 1. 

\citet{Page2008} suggested excluding schools in which all students belong to the same ethnic group  (i.e. ethnically homogeneous schools) because these do not contribute to the estimation of the within-school gap. However, in Colombia, this results in the exclusion of 236,794 students, which constitute 51.6\% of the observations. Therefore, an analysis based on a dataset without ethnically homogeneous schools is unlikely to provide an accurate representation of differences between schools. However, the magnitude of the results presented here may be skewed towards the scores obtained by students in these schools.

\begin{table}[hbt] \centering 
  \caption{Estimation results for the models underlying the decomposition approaches. $M_{ij}$: Dummy variable for minority students. $\overline{M}_{.j}$: School proportion of minority students.} 
  \label{tb:meappcol} 
\small 
\begin{tabular}{@{\extracolsep{5pt}}lD{.}{.}{-3} D{.}{.}{-3} } 
\\[-1.8ex]\hline 
\hline \\[-1.8ex] 
\\[-1.8ex] & \multicolumn{2}{c}{maths} \\ 
\\[-1.8ex] & \multicolumn{1}{c}{(1)} & \multicolumn{1}{c}{(2)}\\ 
\hline \\[-1.8ex] 
 Intercept & 0.031^{***} & 0.046^{***} \\ 
  & (0.002) & (0.002) \\ 
  $M_{ij}$ & -0.459^{***} & -0.072^{***} \\ 
  & (0.006) & (0.010) \\ 
  $\overline{M}_{.j}$ &  & -0.606^{***} \\ 
  &  & (0.012) \\ 
 Observations & \multicolumn{1}{c}{458,947} & \multicolumn{1}{c}{458,947} \\ 
R$^{2}$ & \multicolumn{1}{c}{0.013} & \multicolumn{1}{c}{0.019} \\ 
Adjusted R$^{2}$ & \multicolumn{1}{c}{0.013} & \multicolumn{1}{c}{0.019} \\ 
Residual Std. Error & \multicolumn{1}{c}{0.993} & \multicolumn{1}{c}{0.991} \\ 
F Statistic & \multicolumn{1}{c}{6,149.987$^{***}$} & \multicolumn{1}{c}{4,336.309$^{***}$} \\ 
\hline \\[-1.8ex] 
\textit{Notes:} & \multicolumn{2}{l}{$^{***}$Significant at the 1 percent level.} \\ 
 & \multicolumn{2}{l}{$^{**}$Significant at the 5 percent level.} \\ 
 & \multicolumn{2}{l}{$^{*}$Significant at the 10 percent level.} \\ 
 & \multicolumn{2}{l}{Cluster-robust standard errors.} \\ 
\end{tabular} 
\end{table}

The first model in Table \ref{tb:meappcol} estimates the overall ethnic achievement gap, showing that minority students score on average 0.46 \gls{sd} ($\widehat{\beta}$) lower than their White peers. The second model in this table estimates the within-school gap and the school contextual effect of ethnicity. Accordingly, it shows that minority students score on average 0.07 \gls{sd} ($\widehat{\beta^W}$)  lower than White students attending the same school. Additionally, students attending schools with a ten percentage points higher proportion of minority students score on average 0.06 \gls{sd} ($\frac{\widehat{\beta^{C}}}{10}$) lower, independently of their own ethnicity. The second model in Table \ref{tb:meappcol}  also shows that the average test score of schools that only serve White students is on average 0.68 \gls{sd} {($\widehat{\beta^{S}}=\widehat{\beta^W}+\widehat{\beta^{C}}$)} higher than that of schools only serving minority students.

Additionally, the estimation of the mediation equation  \eqref{eq:mods}, regressing $\overline{M}_{.j}$ on $M_{ij}$, shows that minority students attend schools with 63.8 percentage points ($\widehat{ \omega}$) more minority students than White students. That is, while White students attend schools with an average proportion of minority students of 2.4\% ($\widehat{\gamma}$) , minority students attend schools with an average proportion of minority students of  66.2\% {($\widehat{\gamma}+\widehat{ \omega}$)}.

Section \ref{sx:aglitreview} showed that we can combine this information to decompose the overall ethnic achievement gap. The results of applying these different approaches are presented in Figure \ref{fg:Tbapp4Col}. Under Approach 1, the 0.46 \gls{sd}  overall  gap is decomposed into the 0.07 \gls{sd} within-school gap $\widehat{\beta^W}$ and the 0.39 \gls{sd} effect of segregation $\widehat{\omega}\widehat{\beta^{C}}$, corresponding to the 15.7\% and 84.3\%  of the overall gap, respectively. 

\begin{knitrout}
\definecolor{shadecolor}{rgb}{0.969, 0.969, 0.969}\color{fgcolor}\begin{figure}[htb]

{\centering \includegraphics[width=\textwidth]{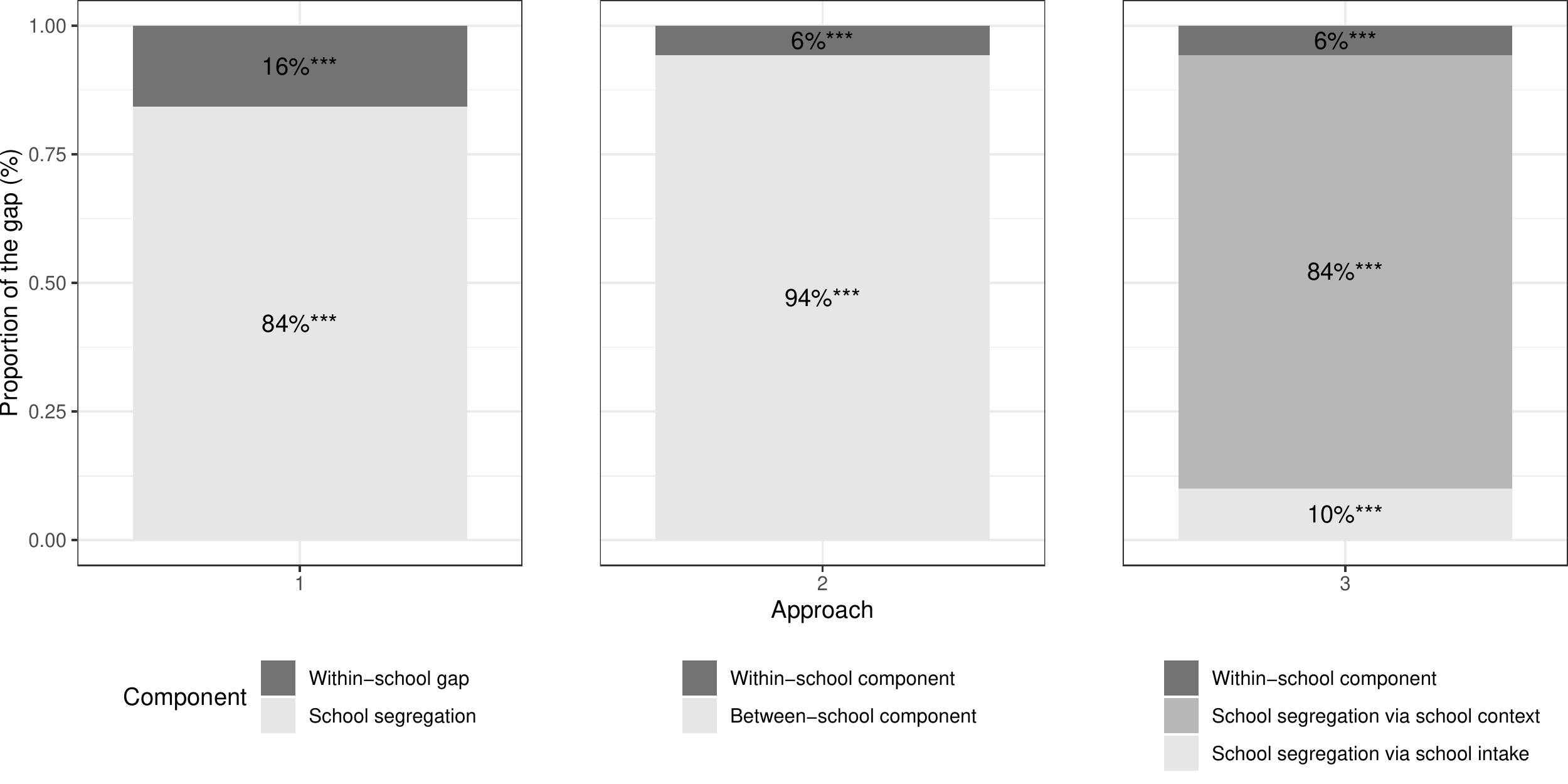} 

}

\caption[White-minority maths achievement gap decomposition under the three different approaches using Colombian data]{White-minority maths achievement gap decomposition under the three different approaches using Colombian data.\\ Note: *** p-value$<$0.001.}\label{fg:Tbapp4Col}
\end{figure}

\end{knitrout}

Approach 2 decomposes the overall gap into the 0.03 \gls{sd} that can be attributed to the within-school gap component $\left(1-\widehat{\omega}\right)\widehat{\beta^W}$ and the 0.43 \gls{sd} that can be attributed to the  between-school gap component $\widehat{\omega}\widehat{\beta^{S}}$, accounting for the  5.7\% and 94.3\%  of the overall gap, respectively. Finally, approach 3 decomposes the gap into 0.03 \gls{sd} attributed to the within-school component of the gap $\left(1-\widehat{\omega}\right)\widehat{\beta^W}$, 0.39 \gls{sd} attributed to the effect of segregation through the contextual effect of ethnicity $\widehat{\omega}\widehat{\beta^{C}}$, and 0.05 \gls{sd} attributed to the effect of segregation through differences in student intake $\widehat{\omega}\widehat{\beta^{W}}$, which correspond to the 5.7\%,  84.3\% and 10\%  of the overall gap, respectively.

In the Colombian context, differences within schools are very small, especially in comparison to those found in the US. For that reason, unlike in the US, in Colombia the conclusion is that most of the ethnic achievement gap is related to school-level processes, regardless of the decomposition approach that we use.

\section{Considering Additional Levels}\label{sx:multlevels}
Following the presentation in section \ref{sx:aglitreview},  this section and section \ref{sx:multgroups} use the mediation analysis framework as a device to extend the current ethnic achievement gap decomposition approaches.  This section focuses on incorporating additional levels while section \ref{sx:multgroups}  considers multiple ethnic groups.

\begin{figure}[hbt]
\begin{center}
\includegraphics[scale=0.65]{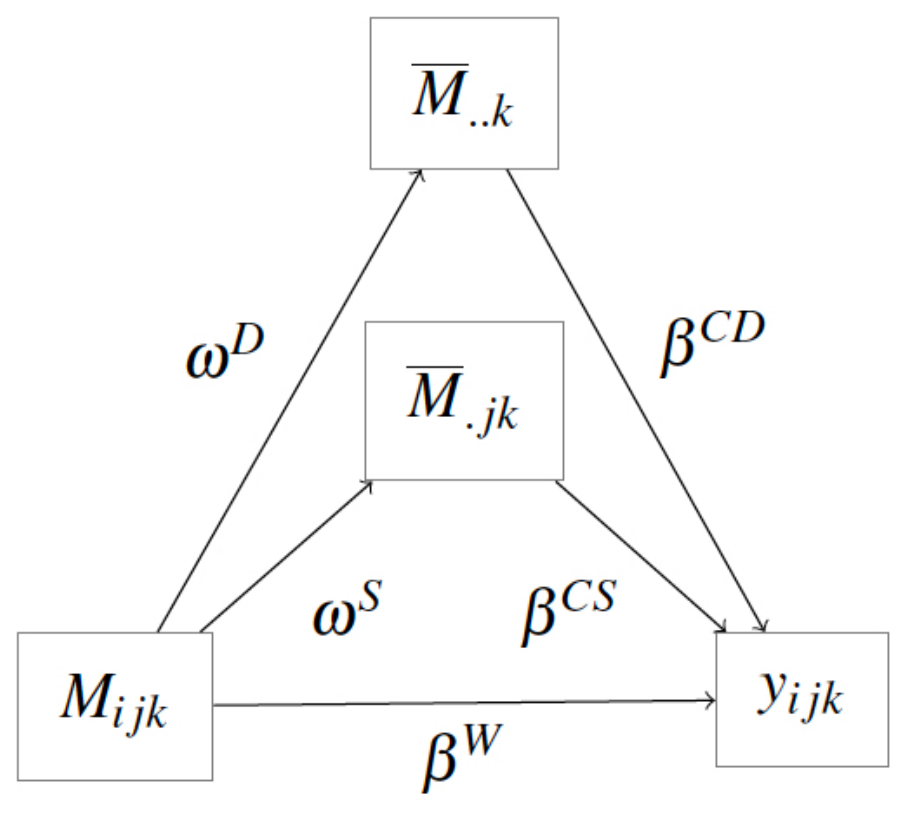}
\caption{Outcome and mediation models for the achievement gap decomposition with an additional level}\label{fg:medmodla}
\end{center}
\end{figure}

To consider an additional level (\acrlong{la}) it is enough to consider a parallel multiple mediation model \citep[Ch.  5]{Hayes2017}  in which the school and \gls{la} proportions of minority students, $\overline{M}_{.jk}$ and $\overline{M}_{..k}$,  mediate the relationship between the ethnicity $M_{ijk}$ and  test scores $y_{ijk}$ of student $i$ attending school $j$ in \gls{la} $k$.  The total effects model in this case is  

\begin{equation}
\begin{matrix}
y_{ijk} = \alpha +  \beta M_{ijk}+{e_y}_{ijk}\\
\\
{e_y}_{ijk} \sim N\left(0,\Omega\right)\\
\end{matrix}
\end{equation}

\noindent which is similar to \eqref{eq:gap} as $\beta$  represents the overall achievement gap between minority and White students.  Figure \ref{fg:medmodla} represents the outcome and the mediation models.

The outcome model is the contextual effect model
\begin{equation}\label{eq:contxla}
y_{ijk}  = \alpha+ \beta^W M_{ijk}+ \beta^{CS} \overline{M}_{.jk} + \beta^{CD} \overline{M}_{..k} +{e_y}_{ijk}
\end{equation}

\noindent where $\beta^W$ is the within-school (and \gls{la}) achievement gap,  $\beta^{CS}$ is the school contextual effect of ethnicity,  conditional on the  \gls{la} ethnic composition and $\beta^{CD}$ is the \gls{la} contextual effect of ethnicity,  conditional on the school ethnic composition.  Thus, students attending schools with a ten percentage points higher proportion of minority students are expected to score $\frac{\beta^{CS}}{10}$ more,  independently of their own ethnicity and \gls{la}'s ethnic composition.  Similarly, students in \glspl{la} with a ten percentage point higher proportion of minority students are expected to score $\frac{\beta^{CD}}{10}$ more,  independently of their ethnicity and school ethnic composition.  This implies that $\beta^{CS}$ is now the school contextual effect of ethnicity within \glspl{la}. 

Since the model includes two mediators, there are two mediation equations
\begin{equation}\label{eq:modla}
\begin{array}{rcl}
\overline{M}_{.jk} & = &\gamma_S + \omega^S M_{ijk} + {e_S}_{.jk}, \forall i \in j,k \\
\overline{M}_{..k} & = &\gamma_D + \omega^{D} M_{ijk} + {e_{D}}_{..k} , \forall i \in k \\
\end{array}
\end{equation}

Again,  \eqref{eq:contxla} and \eqref{eq:modla} are used as devices to decompose the ethnic achievement gap rather than as models with a particular interpretation.  In this case, \eqref{eq:modla} allows us to estimate the between-school segregation index $\omega^S$ and the between-\gls{la} segregation index $\omega^{D}$.  Notice that considering the \gls{la} level does not imply that the between-school segregation index $\omega^S$ is conditional on the \gls{la}, as it is the same as in \eqref{eq:mods}.  That is, $\omega^S$ in this case is not a within-school/between-\gls{la} index.  Instead,  we calculate the school segregation index without considering the \gls{la} to which the schools belong.  Additionally,  the parallel multiple mediation model allows the different mediators ($\overline{M}_{.jk}$ and $\overline{M}_{..k}$) to be correlated but it assumes that  the \gls{la} ethnic composition  $\overline{M}_{..k}$ and academic achievement $y_{ijk}$ are only linked through the \gls{la}  contextual effect of ethnicity $\beta^{CD}$  and not through the school contextual effect of ethnicity $\beta^{CS}$.  The model also assumes that the processes that drive segregation at the school level are different from those at the \gls{la} level.    

In a parallel multiple mediation model \citep{Hayes2017},  given these relationships,  the overall gap $\beta$ can be decomposed as
\begin{equation}\label{eq:mdecl}
\beta= \beta^W + \omega^S\beta^{CS} + \omega^{D}\beta^{CD}
\end{equation}

\noindent where $\beta^W$ would be interpreted as the direct effect and {$\omega^S\beta^{CS}$} and {$\omega^{D}\beta^{CD}$} as specific indirect effects corresponding to the school and \gls{la} ethnic composition,  respectively,  which add up to the  the total indirect effect.  In Approach 1,  this would be equivalent to decomposing the overall gap into the within-school gap $\beta^W$ and the effect of segregation at the school  ({$\omega^S\beta^{CS}$}) and \gls{la} ({$\omega^{D}\beta^{CD}$}) levels.

We know that {$\beta^{CS}=\beta^{BS} - \beta^{W}$},  that is,  the school contextual effect $\beta^{CS}$ is the average  change in school average test scores as the school proportion of minority students increases $\beta^{BS}$, net of the effect of the students' own ethnicity $ \beta^{W}$. Similarly,  {$\beta^{CD}=\beta^{BD}-\beta^{BS}$}, the \gls{la} contextual effect of ethnicity  $\beta^{CD}$ is equivalent to the differences in mean test scores between \glspl{la} with different ethnic composition $\beta^{BD} $ net of the effect of differences between schools  $\beta^{BS}$. Recognising these in \eqref{eq:mdecl} leads to Approach 2, which allows us to write the gap as a combination of the within-school, between-school/within-\gls{la} and between-\gls{la} gaps as

\begin{equation}\label{eq:odecl}
\beta= \left( 1- \omega^{S}\right) \beta^W +\left(\omega^{S}-\omega^{D}\right)\beta^{BS} + \omega^{D}\beta^{BD}
\end{equation}

\noindent where {$\left( 1- \omega^{S}\right) \beta^W $}  is the within-school component of the gap, {$\left(\omega^{S}-\omega^{D}\right)\beta^{BS}$} is the between-school/within \gls{la} component and {$\omega^{D}\beta^{BD}$} is the between \gls{la} component of the gap. 

The third approach further decomposes the between-school and between- \gls{la} components of the gap,  by recognising that  {$\beta^{BS}= \beta^{W} + \beta^{CS} $} and {$\beta^{BD}= \beta^{W} + \beta^{CS} + \beta^{CD} $}. This leads to the six-element decomposition:
 
\begin{equation}\label{eq:reardonml}
\beta= \left(1-\omega^S\right) \beta^W + \left(\omega^S-\omega^D\right)\beta^{W}+ \left(\omega^S-\omega^D\right)\beta^{CS}  + \omega^D\beta^W+ \omega^D\beta^{CS}+ \omega^D\beta^{CD}
\end{equation}

\noindent where {$ \left(1-\omega^S\right) \beta^W$} is the within-school component of the gap, {$\left(\omega^S-\omega^D\right)\beta^{W}$} is the contribution of school segregation through differences in the schools' student intake, {$\left(\omega^S-\omega^D\right)\beta^{CS}$} is the contribution of school segregation through the school contextual effect of ethnicity, {$ \omega^D\beta^W$} is the contribution of \gls{la} segregation through differences in the student intake, {$\omega^D\beta^{CS}$} is the contribution of \gls{la} segregation through differences in the \gls{la}s' school composition and {$\omega^D\beta^{CD}$} is the contribution of \gls{la} segregation through the \gls{la} contextual effect of ethnicity.

The importance of considering additional levels of the education system when applying these three decomposition approaches is likely to depend on the context. The following section illustrates how the implications for policy and practice can be better informed by providing more detailed evidence of the components of the ethnic achievement gap in Colombia. 

\subsection{Application}\label{sx:appml}
As shown in the first model in Table \ref{tb:meappcolml}, when considering additional levels of the school system, the total effect model remains the same as in the simple case shown in Table \ref{tb:meappcol}, which estimates an overall White-minority gap of 0.46 \gls{sd}. In turn, the outcome model (third model in Table \ref{tb:meappcolml}) now includes the \gls{la} proportion of minority students $\overline{M}_{..k}$.

\begin{table}[!hbt] \centering 
  \caption{Estimation results for the models underlying the decomposition approaches. $M_{ijk}$: Dummy variable for minority students. $\overline{M}_{.jk}$: School proportion of minority students.  $\overline{M}_{.jk}$: \gls{la} proportion of minority students.} 
  \label{tb:meappcolml} 
\small 
\begin{tabular}{@{\extracolsep{5pt}}lD{.}{.}{-3} D{.}{.}{-3} D{.}{.}{-3} } 
\\[-1.8ex]\hline 
\hline \\[-1.8ex] 
\\[-1.8ex] & \multicolumn{3}{c}{maths} \\ 
\\[-1.8ex] & \multicolumn{1}{c}{(1)} & \multicolumn{1}{c}{(2)} & \multicolumn{1}{c}{(3)}\\ 
\hline \\[-1.8ex] 
 Intercept & 0.031^{***} & 0.046^{***} & 0.076^{***} \\ 
  & (0.002) & (0.002) & (0.002) \\ 
  $M_{ijk}$ & -0.459^{***} & -0.072^{***} & -0.072^{***} \\ 
  & (0.006) & (0.010) & (0.010) \\ 
  $\overline{M}_{.jk}$ &  & -0.606^{***} & -0.305^{***} \\ 
  &  & (0.012) & (0.013) \\ 
  $\overline{M}_{..k}$ &  &  & -0.747^{***} \\ 
  &  &  & (0.014) \\ 
 Observations & \multicolumn{1}{c}{458,947} & \multicolumn{1}{c}{458,947} & \multicolumn{1}{c}{458,947} \\ 
R$^{2}$ & \multicolumn{1}{c}{0.013} & \multicolumn{1}{c}{0.019} & \multicolumn{1}{c}{0.024} \\ 
Adjusted R$^{2}$ & \multicolumn{1}{c}{0.013} & \multicolumn{1}{c}{0.019} & \multicolumn{1}{c}{0.024} \\ 
Residual Std. Error & \multicolumn{1}{c}{0.993} & \multicolumn{1}{c}{0.991} & \multicolumn{1}{c}{0.988} \\ 
F Statistic & \multicolumn{1}{c}{6,149.987$^{***}$} & \multicolumn{1}{c}{4,336.309$^{***}$} & \multicolumn{1}{c}{3,748.908$^{***}$} \\ 
\hline \\[-1.8ex] 
\textit{Notes:} & \multicolumn{3}{l}{$^{***}$Significant at the 1 percent level.} \\ 
 & \multicolumn{3}{l}{$^{**}$Significant at the 5 percent level.} \\ 
 & \multicolumn{3}{l}{$^{*}$Significant at the 10 percent level.} \\ 
 & \multicolumn{3}{l}{Cluster-robust standard errors.} \\ 
\end{tabular} 
\end{table}

As shown in Table \ref{tb:meappcolml}, including $\overline{M}_{..k}$ into the model leaves the 0.07 \gls{sd} estimate of the within-school ethnic achievement gap  $\widehat{\beta^W}$ unaffected. In turn, it isolates the \gls{la} contextual effect of ethnicity from the original estimate of the school contextual effect of ethnicity (second model in the table). Hence, it is now estimated that within \glspl{la}, 
students attending schools with a ten percentage points higher proportion of minority students score on average 0.03 \gls{sd} ($\frac{\widehat{\beta^{CS}}}{10}$) lower, independently of their own ethnicity. Additionally, students attending \glspl{la} with a ten percentage points higher proportion of minority students score on average 0.07 \gls{sd} ($\frac{\widehat{\beta^{CD}}}{10}$) lower, independently of their own ethnicity and the school they attend.

The third model in Table \ref{tb:meappcolml}  also shows that, within \glspl{la}, schools that only serve White students score on average 0.38 \gls{sd} {($\widehat{\beta^{BS}}=\widehat{\beta^W}+\widehat{\beta^{CS}}$)} higher than schools only serving minority students.  Additionally, \glspl{la} that only serve White students score on average 1.12 \gls{sd} {($\widehat{\beta^{BS}}=\widehat{\beta^W}+\widehat{\beta^{CS}}+\widehat{\beta^{CD}}$)} higher than those that only serve minority students. In our data, this is an hypothetical comparison, as there are no \glspl{la} with a  proportion of minority students of 100\%.

The estimation of mediation equations \eqref{eq:modla} shows that minority students attend schools and \glspl{la} with 63.8 and 25.8 percentage points more minority students than White students ($\widehat{ \omega^S}$ and $\widehat{ \omega^D}$), respectively. As  discussed in the previous section, the different combinations of these parameters lead to the three different decompositions presented in Figure \ref{fg:Tbapp4ColML}.

\begin{knitrout}
\definecolor{shadecolor}{rgb}{0.969, 0.969, 0.969}\color{fgcolor}\begin{figure}[htb]

{\centering \includegraphics[width=\textwidth]{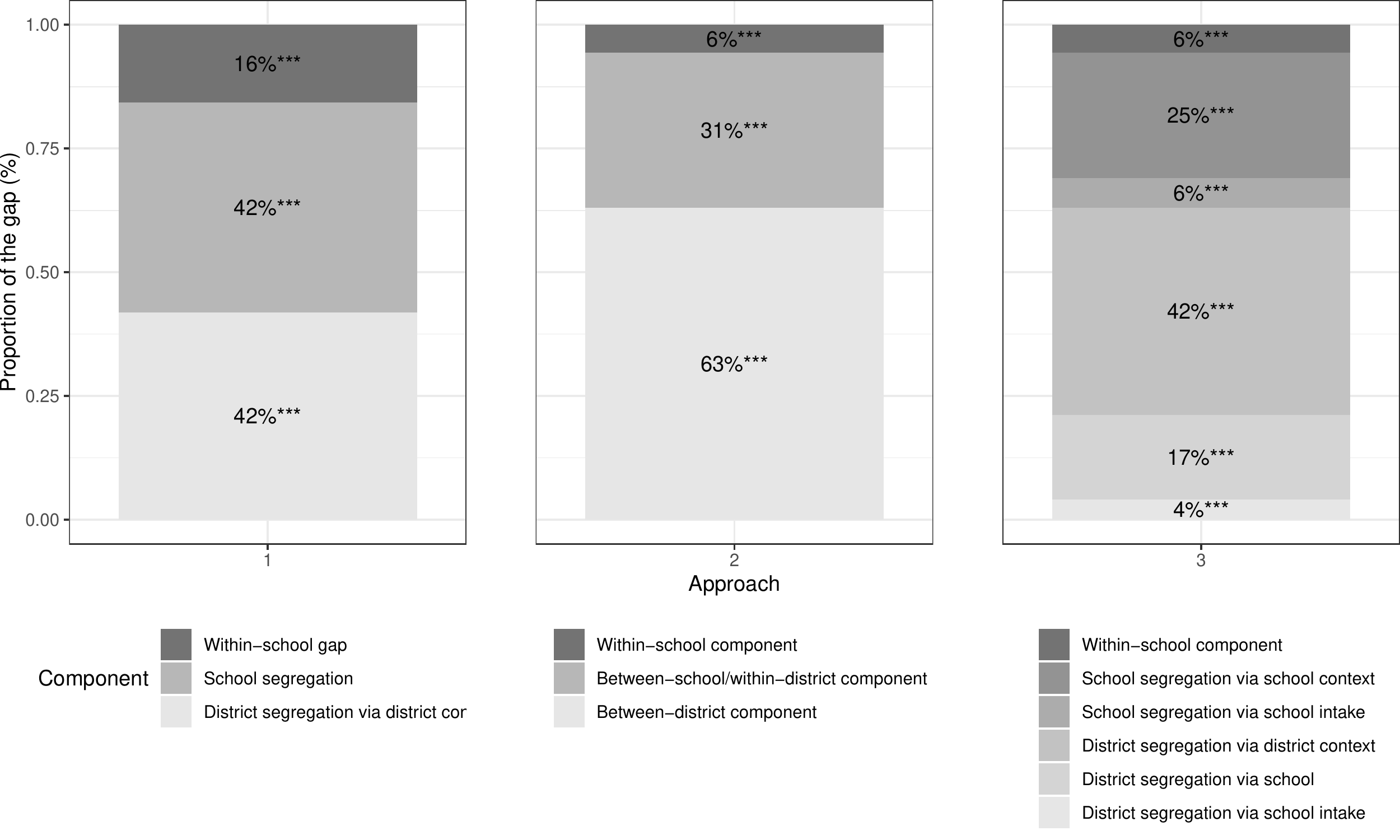} 

}

\caption[White-minority maths achievement gap decomposition under the three different approaches considering  {\glspl{la}}]{White-minority maths achievement gap decomposition under the three different approaches considering  {\glspl{la}}. \\ Note: *** p-value$<$0.001.}\label{fg:Tbapp4ColML}
\end{figure}

\end{knitrout}

Under Approach 1, the 0.46 \gls{sd}  overall  gap is decomposed into the 0.07 \gls{sd} within-school gap $\widehat{\beta^W}$,  the 0.19 \gls{sd}  effect of school segregation $\widehat{\omega^S}\widehat{\beta^{CS}}$ and the 0.19 \gls{sd} effect of \gls{la} segregation $\widehat{\omega^D}\widehat{\beta^{CD}}$. These correspond to  15.7\% , 42.4\%  and 41.9\%  of the overall gap, respectively. Approach 2 decomposes the overall gap into the 0.03 \gls{sd} that can be attributed to the within-school gap component $\left(1-\widehat{\omega^S}\right)\widehat{\beta^W}$, the 0.14 \gls{sd} that can be attributed to the  between-school gap component $\widehat{\omega^S}\widehat{\beta^{BS}}$ and  the 0.29 \gls{sd} that can be attributed to the  between-\gls{la} gap component $\widehat{\omega^D}\widehat{\beta^{BD}}$. These elements account for the  5.7\%,  31.3\% and 63\%  of the overall gap, respectively.

Finally, approach 3 decomposes the gap into six different components. First, 0.03 \gls{sd} (5.7\%) attributed to the within-school component of the gap {$\left(1-\widehat{\omega^S}\right)\widehat{\beta^W}$}; second, 0.12 \gls{sd} (25.3\%) attributed to the effect of school segregation through the school contextual effect of ethnicity {$\left(\widehat{\omega^S}-\widehat{\omega^D}\right)\widehat{\beta^{W}}$}; third,  0.03 \gls{sd} (6\%) attributed to the effect of school segregation through differences in student intake {$\left(\widehat{\omega^S}-\widehat{\omega^D}\right)\widehat{\beta^{W}}$}; fourth, 0.02 \gls{sd} (4.1\%) attributed to \gls{la} segregation through differences in the schools' student intake {$\widehat{\omega^D}\widehat{\beta^W}$} ; fifth,   0.08 \gls{sd} (17.1\%) attributed to \gls{la} segregation through differences in the districts' school composition ($ \widehat{\omega^D}\widehat{\beta^{CS}}$); finally, {0.19 \gls{sd}} {(41.9\%)} attributed to \gls{la} segregation through differences in the districts' ethnic composition ($ \widehat{\omega^D}\widehat{\beta^{CD}}$).

These three decomposition approaches provide different insights and levels of granularity about the role of various levels of the school system in the appearance of the White-minority achievement gap in Colombia. Nonetheless, all of them draw attention to the importance of differences between \glspl{la}, with  implications for research and policy. An analysis based on the original decomposition approaches would lead to recommendations focusing on schools, according to the results presented in section \ref{sx:agdata}. Such recommendations commonly focus on school improvement, via teacher training or upgrades in school infrastructure \citep[e.g. ][]{Fryer2006a,Fryer2004a}. 

The results in this section show that one important school characteristic that contributes to the ethnic achievement gap is the \gls{la} where the school is located. Hence, policies tackling inequality among \glspl{la} are as (if not more) important for narrowing the White-minority gap in Colombia. Therefore, extending the current decomposition approaches to incorporate the role of \glspl{la} leads to better informed policy recommendations.

\section{Considering Multiple Ethnic Groups}\label{sx:multgroups}

So far, we have assumed that the gaps between all minority groups and White students are the same and that the extent to which the gap can be attributed to differences between students, schools and \glspl{la} is the same for all ethnic groups. These assumptions are probably incorrect. Therefore an alternative decomposition that allows for different gaps and components for each ethnic group may be more attractive. We explore this extension in this section. This and the extension discussed in section \ref{sx:multlevels} can be combined to decompose the achievement gaps of multiple ethnic groups while considering additional levels of the school system, as shown in  appendix \ref{ax:ext}. However, to facilitate the explanation, this section limits the presentation to the two-level case.

Again,  mediation analysis can be used as a device to extend the current gap decomposition approaches to consider multiple ethnic groups.  This time,  it is enough to combine a mediation model with a multicategorical independent variable \citep{Hayes2014S} and a parallel multiple mediation model \citep{Hayes2017}.  For our illustration,  we compare three possible ethnic categories with the White category, which is used as a reference: Afrocolombian $A_{ij}$, Indigenous $I_{ij}$, and other minorities $O_{ij}$. Here, each ethnic group is represented by a dummy variable that equals one when a student belongs to that ethnic group and 0 otherwise. All dummy variables are independent variables, and each one is associated with a unique gap between students of that ethnic group and White students.  Should the comparison of interest be between any other two groups (e.g. Afrocolombian versus Indigenous students), it is enough to change the reference category (e.g. to Afrocolombian students). 

\begin{figure}[hbt]
\begin{center}
\includegraphics[scale=0.65]{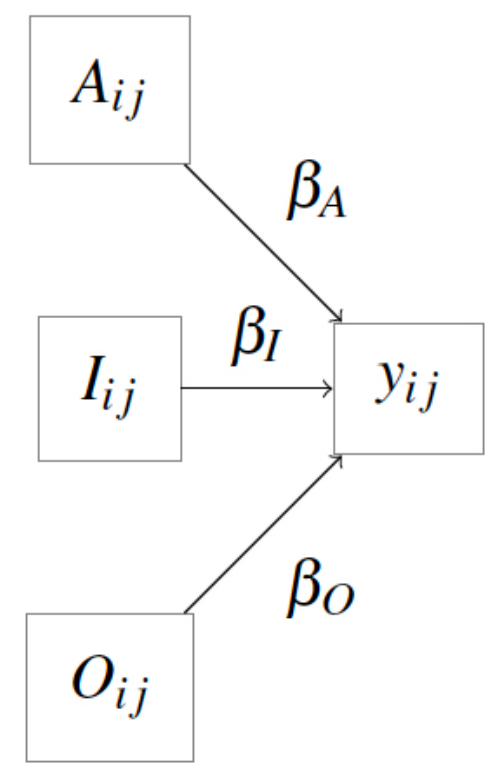}
\caption{Total effect model for the decomposition of the gaps for multiple ethnic minority groups}\label{fg:totmedgroups}
\end{center}
\end{figure}

The total effect model, depicted in Figure \ref{fg:totmedgroups}, therefore is
\begin{equation}
y_{ij}  = \alpha + \beta_A A_{ij}+ \beta_I I_{ij}+ \beta_O O_{ij}+{e_y}_{ij}
\end{equation}

\noindent where $\beta_A$, $\beta_I$ and $\beta_O$ represent the gap in test scores between White and Afrocolombian, Indigenous, other ethnic minorities, respectively.

\begin{figure}[hbt]
\begin{center}
\includegraphics[scale=0.35]{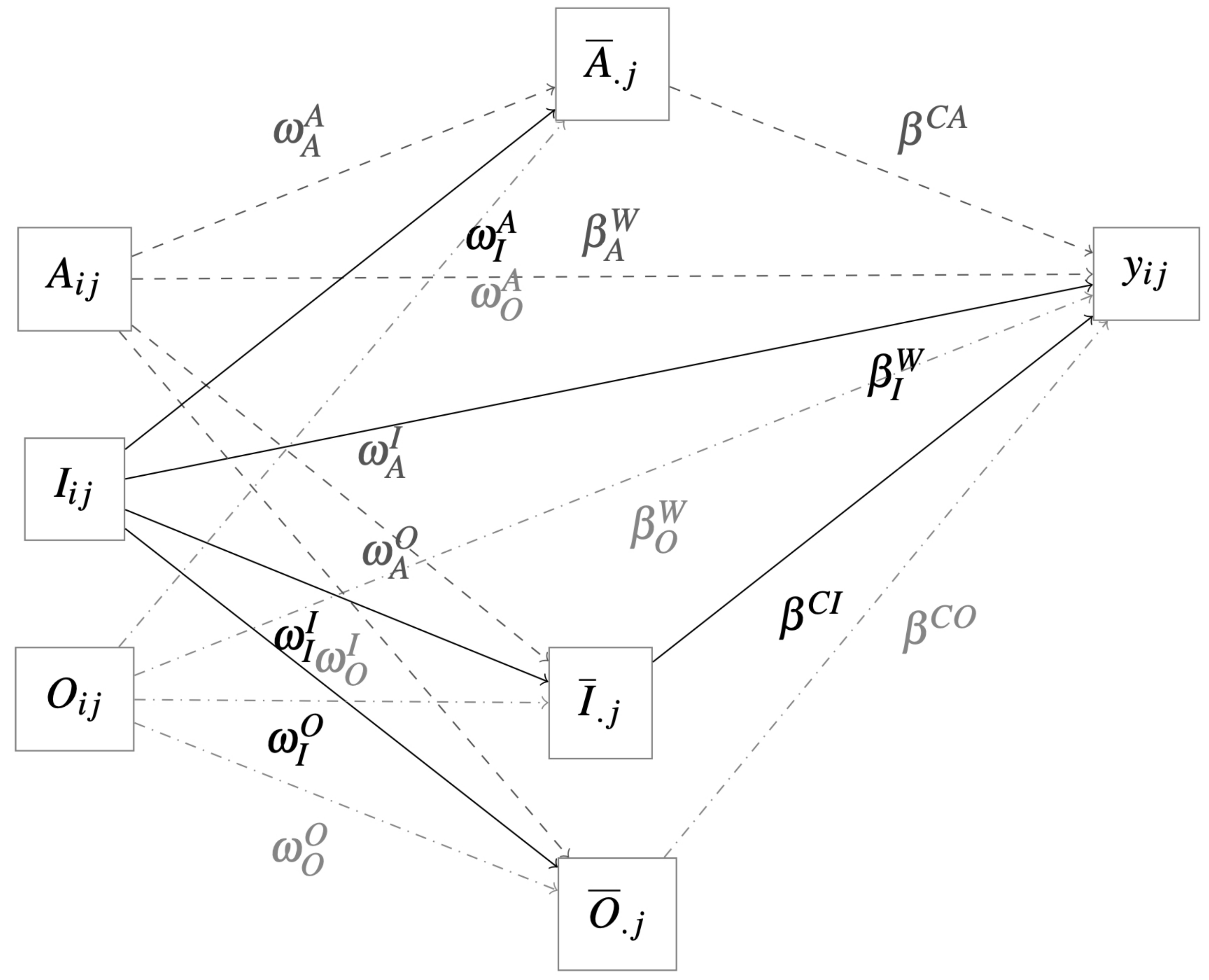}
\caption{Outcome and mediation models for the decomposition of the gaps for different ethnic groups}\label{fg:outmedgroups}
\end{center}
\end{figure}

The outcome and mediation models are represented in Figure \ref{fg:outmedgroups}, which shows that besides having an explanatory variable for each minority group, there is also one mediator for each of them \citep{Hayes2014S}.  These mediators are the school proportion of students belonging to each of the minority groups\footnote{Should the comparison of interest be with respect to other ethnic group (e.g. Afrocolombians), the mediators would be the school proportions of students in the remaining categories (e.g. White, Indigenous and other minorities).}, $\overline{A}_{.j}$, $\overline{I}_{.j}$ and $\overline{O}_{.j}$.  Consequently, the outcome model corresponds to the contextual effect model

\begin{equation}\label{eq:contxg}
y_{ij}  = \alpha + \beta_A^{W} A_{ij}+ \beta_I^{W} I_{ij}+ \beta_O^{W} O_{ij}+ \beta^{CA}\overline{A}_{.j} + \beta^{CI}\overline{I}_{.j} + \beta^{CO}\overline{O}_{.j}+{e_y}_{ij}
\end{equation}

\noindent where $\beta_A^{W}$, $\beta_I^{W}$ and $\beta_O^{W}$ are the within-school achievement gaps for each group and $\beta^{CA}$, $\beta^{CI}$ and $\beta^{CO}$ are the school contextual effects of each minority group, conditional on the school proportion of other ethnic minorities. 

Following \citet{Hayes2017}, there is one mediation model for each mediator, and thus for each ethnic minority group, and  each mediator is a function of the school ethnic composition in terms of all minority groups. As in section \ref{sx:multlevels}, this is only a device to decompose the achievement gap for each minority group, which results in the models 
\begin{equation}\label{eq:modg}
\begin{array}{rcl}
\overline{A}_{.j} &= &\gamma_A + \omega_A^{A} A_{ij} + \omega_I^{A} I_{ij} + \omega_O^{A} O_{ij} + {e_{SA}}_{.j} , \forall i \in j \\
\overline{I}_{.j} &= &\gamma_I + \omega_A^{I} A_{ij} + \omega_I^{I} I_{ij} + \omega_O^{I} O_{ij} + {e_{SI}}_{.j}, \forall i \in j \\
\overline{O}_{.j} &= &\gamma_O + \omega_A^{O} A_{ij} + \omega_I^{O} I_{ij} + \omega_O^{O} O_{ij} + {e_{SO}}_{.j}, \forall i \in j \\
\end{array}
\end{equation}

Again the mediation equations provide estimates of indicators of segregation, given by the differences in the mean school proportion of students of each minority group for each minority and White students. For instance, $\omega_A^{I}$ is the difference in the mean school proportion of Indigenous students between Afrocolombian and White students; if {$\omega_A^{I}=0$}, White and Afrocolombian students attend schools with the same average proportion of Indigenous students. 

Therefore, following \citet{Hayes2014S,Hayes2017}, each of the gaps can be decomposed into their relative direct and (specific and total) indirect effects analogously to \eqref{eq:mdecl} in  section \ref{sx:multlevels}. There is a relative direct effect for each ethnic minority, which corresponds to the within-school achievement gap.  Each of the mediators contributes to the  specific relative indirect effects  which add up to the total relative indirect effects for each ethnic group.  Importantly,  \eqref{eq:modg} implies that the achievement gap of each ethnic group is not only a function of its own contextual effect but also a function of the contextual effects of other minority groups. For example, the overall ethnic achievement gap for Afrocolombians can be decomposed into the direct effect $\beta_A^{W}$  and the indirect effect {$\omega_A^{A} \beta^{CA} + \omega_A^{I}\beta^{CI} + \omega_A^{O}\beta^{CO}$}. In general, 

\begin{equation}\label{eq:mgmeddec}
\begin{array}{rcl}
\beta_A &= &\beta_A^W + \omega_A^{A} \beta^{CA} + \omega_A^{I} \beta^{CI}+ \omega_A^{O}\beta^{CO}\\
\beta_I &= & \beta_I^W + \omega_I^{A} \beta^{CA} + \omega_I^{I} \beta^{CI} + \omega_I^{O} \beta^{CO}\\
\beta_O &= & \beta_O^W + \omega_O^{A} \beta^{CA} + \omega_O^{I} \beta^{CI}+ \omega_O^{O} \beta^{CO}\\
\end{array}
\end{equation}

This shows that the multiple-level decomposition of the overall ethnic achievement gaps for multiple ethnic groups is not possible without incorporating the contextual effects of the other minority groups, which are weighted according to their relative (to that of White students) exposure  to other ethnic groups. For example, the overall achievement gap for Afrocolombian students does not only depend on the within-school gap and the school contextual effect of Afrocolombian students but also on the contextual effects of  Indigenous and other minority students. The empirical attempts to incorporate several groups into the analysis  \citep{Quinn2015a,Dustmann2008,Page2008} have not recognised the role of the remaining ethnic groups as part of a school-level element of the gap. The importance of this omission depends on how segregated minority groups are with respect to each other and how strong their contextual effects are.

As before,  the school proportions of each  ethnic minority (the mediators) can be correlated \citep{Hayes2017},  but the model implies that they are linked to academic achievement only through the contextual effect of each ethnicity (as opposed to, for example, the school proportion of Afrocolombian students $\overline{A}_{.j}$  being linked to academic achievement through the contextual effect of Indigenous students  $\beta^{CI}$).  Alternatively, performing pair-wise comparisons between white and each minority group (e.g. Afrocolombian) potentially brings additional distortions into the analysis. Performing the analysis by restricting the sample to include only two groups (e.g. White and Afrocolombian) of students incorporates measurement error in the estimated school proportion of minority students, by reducing the total number of students within each school, and thus into the decomposition analysis. The alternative of focusing only on schools that are exclusively attended by each pair of ethnic groups (e.g. White and Afrocolombian) might exclude a large proportion of schools and therefore ignore a substantial source of variation when estimating the school contextual effects.

Under Approach 1, \eqref{eq:mgmeddec} is equivalent to decomposing the overall gap into the within school gap {(e.g. $\beta_A^{W}$)} and the effect of segregation with respect to the White group  (e.g. {$\omega_A^{A} \beta^{CA} + \omega_A^{I}\beta^{CI} + \omega_A^{O}\beta^{CO}$}), which results on a different exposure to the contextual effects of minority groups, in comparison with White students.

Recognising that the between-school gap for each ethnicity depends on the within-school gap and contextual effect of such ethnicity (e.g., {$\beta^{SA}=\beta_A^W+\beta^{CA}$)} in \eqref{eq:mgmeddec} leads to Approach 2:

\begin{equation}\label{eq:odecg}
\begin{array}{rcl}
\beta_A &=& \left(1-\omega_A^{A}\right) \beta_A^W + \omega_A^{A}\beta^{SA} + \omega_A^{I}\beta^{CI} + \omega_A^{O}\beta^{CO}   \\
\beta_I &=& \left(1-\omega_I^{I}\right) \beta_I^W + \omega_I^{I}\beta^{SI} + \omega_I^{A}\beta^{CA} + \omega_I^{O}\beta^{CO}   \\
\beta_O &=& \left(1-\omega_O^{O}\right) \beta_O^W + \omega_O^{O}\beta^{SO} + \omega_O^{A}\beta^{CA}+ \omega_O^{I}\beta^{CI}   \\
\end{array}
\end{equation}
  
\noindent where, for example,  {$\left(1-\omega_A^{A}\right) \beta_A^W$} is the within-school component of the overall Afrocolombian-White gap, {$ \omega_A^{A}\beta^{SA}$} is the component of the gap that can be attributed to the between-school gap for Afrocolombian students, and {$\omega_A^{I}\beta^{CI} + \omega_A^{O}\beta^{CO}$} is the effect of a differential exposure to other ethnic minority groups. 

Intuitively, the overall Afrocolombian-White between-school gap $\beta^{SA}$  is the result of comparing schools with an increasing proportion of Afrocolombian students, while keeping everything else constant. Nonetheless, schools with the same proportion of Afrocolombian students may differ in their proportion of Indigenous and other ethnic minority students. Hence, when decomposing the gap we need to acknowledge these additional differences by including the components of the other ethnic minority groups. 

Approach 3 implies further decomposing the between-school component of the gap, recognising that {$\beta^{SA}=\beta_A^W+\beta^{CA}$}, {$\beta^{SI}=\beta_I^W+\beta^{CI}$} and {$\beta^{SO}=\beta_O^W+\beta^{CO}$}, which leads to

\begin{equation}\label{eq:odecgrd}
\begin{array}{rcl}
\beta_A &=& \left(1-\omega_A^{A}\right) \beta_A^W + \omega_A^{A}\beta_A^{W} + \omega_A^{A}\beta^{CA} + \omega_A^{I}\beta^{CI} + \omega_A^{O}\beta^{CO}   \\
\beta_I &=& \left(1-\omega_I^{I}\right) \beta_I^W + \omega_I^{I}\beta_I^{W}+ \omega_I^{I}\beta^{CI} + \omega_I^{A}\beta^{CA} + \omega_I^{O}\beta^{CO}   \\
\beta_O &=& \left(1-\omega_O^{O}\right) \beta_O^W + \omega_O^{O}\beta_O^{W}+ \omega_O^{O}\beta^{CO} + \omega_O^{A}\beta^{CA}+ \omega_O^{I}\beta^{CI}   \\
\end{array}
\end{equation}

\noindent where, taking the Afrocolombian-White achievement gap as an example, {$\left(1-\omega_A^{A}\right) \beta_A^W$} is the part of the gap that can be attributed to the within-school gap, {$\omega_A^{A}\beta_A^{W}$} is the component of the gap that can be attributed to segregation through differences in the schools' student intake, and {$\omega_A^{A}\beta^{CA} + \omega_A^{I}\beta^{CI} + \omega_A^{O}\beta^{CO}$} is the effect of segregation through the contextual effect of ethnicity, which can be further decomposed to recognise the contribution of a differential exposure to each minority group.

Again, depending on the context, incorporating multiple ethnic groups in the decomposition may improve the relevance of evidence for policy and future research. The next section shows that this is indeed the case for Colombia.

\subsection{Application}\label{sx:appmg}
As discussed in section \ref{sx:multgroups}, the multiple-ethnic-group decomposition uses the parameters of the total and outcome models presented in Table \ref{tb:meappcolmg}.  As shown, the 0.59 \gls{sd} White-Afrocolombian achievement gap $\widehat{\beta_A}$ is wider than the 0.52 \gls{sd} White-Indigenous $\widehat{\beta_I}$ and the 0.14 \gls{sd} White-Other $\widehat{\beta_O}$ achievement gaps. These differences in the overall achievement gaps underscore the importance of distancing from the aggregated measure of the gap presented in sections \ref{sx:agdata} and \ref{sx:appml} to consider the richer information of multiple ethnic minorities.  


\begin{table}[!h] \centering 
  \caption{Estimation results for the models underlying the decomposition approaches. $A_{ij}$, $I_{ij}$, $O_{ij}$: Dummy variables for Afrocolombian, Indigenous and other minority students.  $\overline{A}_{.j}$, $\overline{I}_{.j}$ and $\overline{O}_{.j}$: School proportion of Afrocolombian, Indigenous and other minority students.} 
  \label{tb:meappcolmg} 
\small 
\begin{tabular}{@{\extracolsep{5pt}}lD{.}{.}{-3} D{.}{.}{-3} } 
\\[-1.8ex]\hline 
\hline \\[-1.8ex] 
\\[-1.8ex] & \multicolumn{2}{c}{maths} \\ 
\\[-1.8ex] & \multicolumn{1}{c}{(1)} & \multicolumn{1}{c}{(2)}\\ 
\hline \\[-1.8ex] 
 Intercept & 0.031^{***} & 0.043^{***} \\ 
  & (0.002) & (0.002) \\ 
  $A_{ij}$ & -0.592^{***} & -0.089^{***} \\ 
  & (0.008) & (0.015) \\ 
  $I_{ij}$ & -0.518^{***} & -0.112^{***} \\ 
  & (0.011) & (0.017) \\ 
  $O_{ij}$ & -0.142^{***} & -0.026^{*} \\ 
  & (0.011) & (0.015) \\ 
  $\overline{A}_{.j}$ &  & -0.707^{***} \\ 
  &  & (0.017) \\ 
  $\overline{I}_{.j}$ &  & -0.669^{***} \\ 
  &  & (0.022) \\ 
  $\overline{O}_{.j}$ &  & -0.172^{***} \\ 
  &  & (0.022) \\ 
 Observations & \multicolumn{1}{c}{458,947} & \multicolumn{1}{c}{458,947} \\ 
R$^{2}$ & \multicolumn{1}{c}{0.016} & \multicolumn{1}{c}{0.021} \\ 
Adjusted R$^{2}$ & \multicolumn{1}{c}{0.016} & \multicolumn{1}{c}{0.021} \\ 
Residual Std. Error & \multicolumn{1}{c}{0.992} & \multicolumn{1}{c}{0.990} \\ 
F Statistic & \multicolumn{1}{c}{2,420.402$^{***}$} & \multicolumn{1}{c}{1,609.472$^{***}$} \\ 
\hline \\[-1.8ex] 
\textit{Notes:} & \multicolumn{2}{l}{$^{***}$Significant at the 1 percent level.} \\ 
 & \multicolumn{2}{l}{$^{**}$Significant at the 5 percent level.} \\ 
 & \multicolumn{2}{l}{$^{*}$Significant at the 10 percent level.} \\ 
 & \multicolumn{2}{l}{Cluster-robust standard errors.} \\ 
\end{tabular} 
\end{table}

Table  \ref{tb:meappcolmg} also shows that within schools, Afrocolombian, Indigenous and other minority groups score  0.09 \gls{sd}  ($\widehat{\beta_A^W}$), 0.11 \gls{sd} ($\widehat{\beta_I^W}$) and 0.03 \gls{sd} ($\widehat{\beta_O^W}$) lower than their White peers. Additionally, a ten percentage-point increase in the proportion of Afrocolombian, Indigenous and other minority students is linked to an average decrease of 0.71 \gls{sd} ($\frac{\widehat{\beta^{CA}}}{10}$),  0.67 \gls{sd} ($\frac{\widehat{\beta^{CI}}}{10}$) and  0.17 \gls{sd} ($\frac{\widehat{\beta^{CO}}}{10}$)   in the students' maths test scores, repectively, independently of their own ethnicity. 

As discussed above, considering multiple ethnic groups requires considering the differences in exposure to them. These differences are displayed in Table \ref{tb:wMG}, which shows that minority groups are more exposed to other students of the same ethnic minority, in comparison to White students. For example, Afrocolombian students attend schools with an average proportion of Afrocolombian students 0.7 percentage points ($\widehat{\omega^A_A}$) higher than the schools White students attend. In turn, the exposure to other minority groups is similar to that of White students. For example, Afrocolombian students attend schools with the same average proportion of Indigenous students as the schools that White students attend ($\widehat{\omega^I_A}=0$).

\begin{table}[htb]
\centering
\caption{School segregation indices for all ethnic minority groups} 
\label{tb:wMG}
\begingroup\small
\begin{tabular}{crrr}
  \hline
Group & \% Afro & \% Indigenous & \% Other \\ 
  \hline
Afro & 0.70 & 0.00 & 0.02 \\ 
  Indigenous & 0.01 & 0.59 & 0.02 \\ 
  Other & 0.04 & 0.02 & 0.43 \\ 
   \hline
\end{tabular}
\endgroup
\end{table}

Each of the three decomposition approaches studied here result from the combination of these parameters. These results are shown in Figure \ref{fg:Tbapp4ColMG}.  Taking the White-Afrocolombian achievement gap as an example, Approach 1 attributes 0.09 \gls{sd} (15.1\%) 
of the White-Afrocolombian achievement gap to the within-school gap {$\widehat{\beta_A^W}$}  and the remainder of the gap {$\widehat{\omega_A^{A}}\widehat{ \beta^{CA}} + \widehat{\omega_A^{I}} \widehat{\beta^{CI}}+ \widehat{\omega_A^{O}}\widehat{\beta^{CO}}$} to the effect of a differential exposure to other minority groups.

\begin{knitrout}
\definecolor{shadecolor}{rgb}{0.969, 0.969, 0.969}\color{fgcolor}\begin{figure}[htb]

{\centering \includegraphics[width=\textwidth]{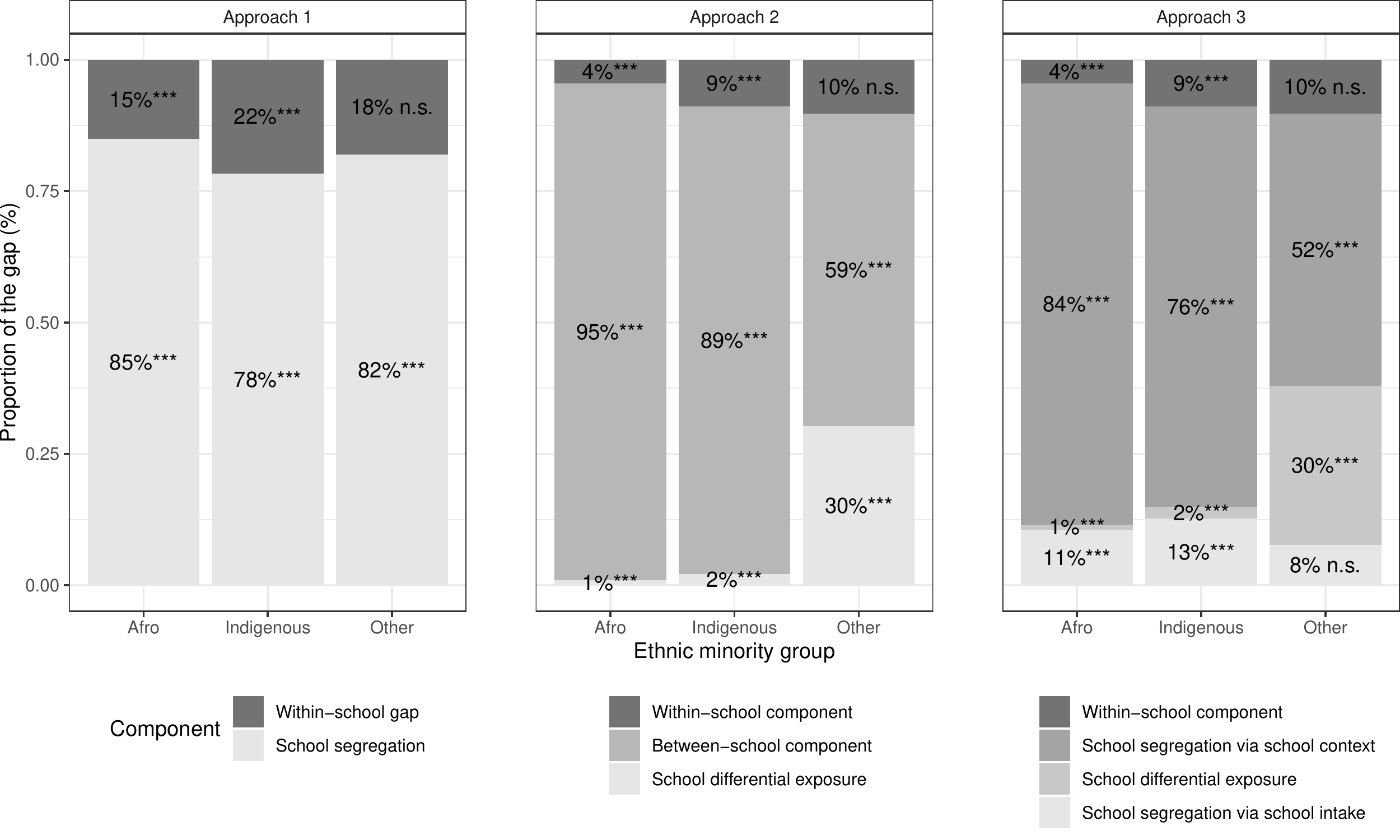} 

}

\caption[White-minority maths achievement gap decomposition under the three different approaches considering multiple ethnic minorities]{White-minority maths achievement gap decomposition under the three different approaches considering multiple ethnic minorities. \\ Note: ***: p-value$<$0.001; n.s: Not statistically significant.}\label{fg:Tbapp4ColMG}
\end{figure}

\end{knitrout}

Approaches 2 and 3 attribute  0.03 \gls{sd} (4.5\%) of the White-Afrocolombian gap to the within-school component of the gap {$\left(1-\widehat{\omega_A^{A}}\right)\widehat{\beta_A^W}$} and { 0.01 \gls{sd}} (1\%) 
of this gap to the effect of a differential exposure to other minority groups  $\widehat{\omega_A^{I}}\widehat{\beta^{CI}} + \widehat{\omega_A^{O}}\widehat{\beta^{CO}}$.  
The difference between these two approaches is that  Approach 2 attributes the remaining  0.56 \gls{sd} (94.5\%) 
of the gap to the between-school component of the gap $\widehat{\omega_A^{A}}\widehat{\beta^{SA}}$.  In turn,  Approach 3 provides two additional components:   0.5 \gls{sd} (83.9\%) for the effect of segregation via differential exposure to the contextual effect of same ethnic group $\widehat{\omega_A^{A}}\widehat{ \beta^{CA}}$ and 0.06 \gls{sd} (10.6\%)  for the effect of segregation via differential student intake $\widehat{\omega_A^{A}}\widehat{ \beta^{WA}}$ .
 
Alternative decompositions can be accommodated. For example, if it is relevant to know which minority group contributes the most to the ethnic achievement gap, the school-level component of the gap can be further decomposed to analyse the individual contributions of the contextual effects of each minority group.   


\section{On Serial Multiple Mediation Models}\label{sx:serialmed}
As discussed in sections \ref{sx:multlevels} and \ref{sx:multgroups},  the extensions to the three existing gap decomposition approaches to consider multiple levels and multiple ethnic groups are based on a parallel multiple mediation model.  An alternative model that considers multiple mediations is the serial multiple mediation model. This section explains why the extensions presented in sections \ref{sx:multlevels} and \ref{sx:multgroups} are not based on a serial multiple mediation model.  We illustrate the discussion using the extension to include multiple levels, but a similar argument can be made about the extension to consider multiple ethnic groups.

\begin{figure}[hbt]
\begin{center}
\begin{tabular}{cc}
\includegraphics[scale=0.5]{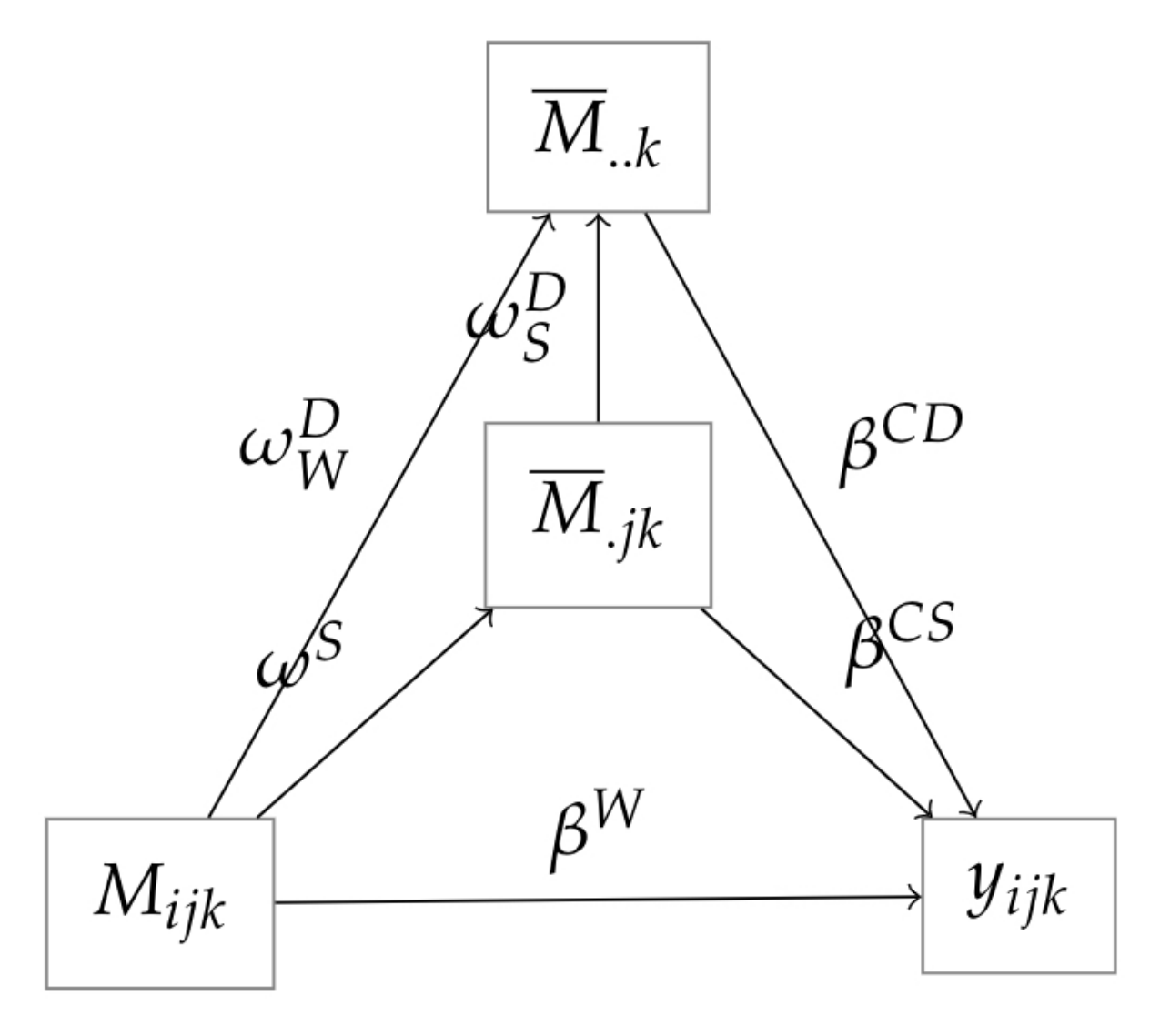} & \includegraphics[scale=0.5]{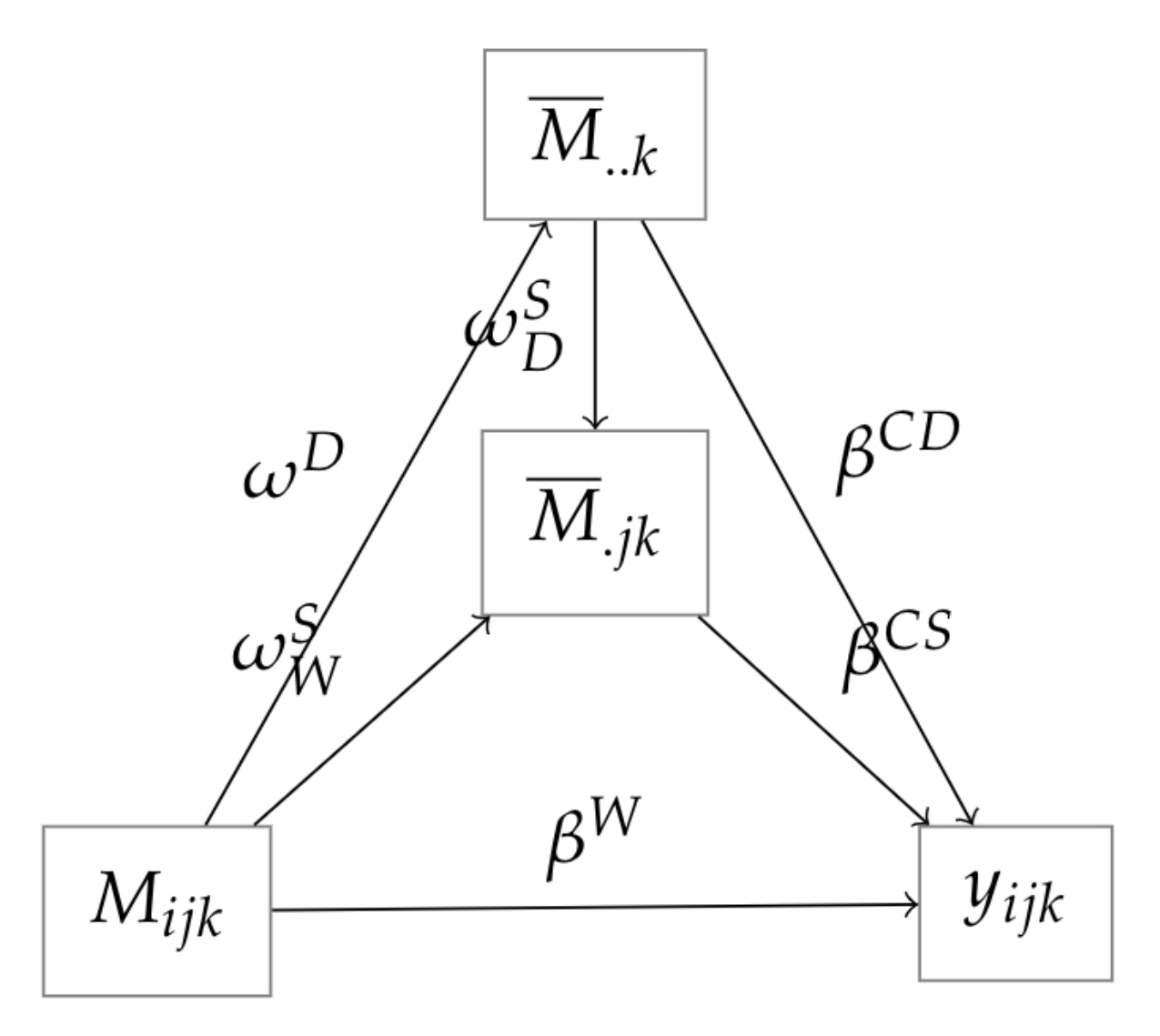} \\
(i) & (ii) \\
\end{tabular}

\caption{Outcome and mediation models for the achievement gap decomposition with an additional level using two alternative serial multiple mediation models}\label{fg:medserial}
\end{center}
\end{figure}

The parallel multiple mediation model assumes that school segregation and \gls{la} segregation are correlated but separate processes.   Alternatively,  serial multiple mediation models could be considered.  In   Figure \ref{fg:medserial},  a new path is added between the school proportion of minority students $\overline{M}_{.jk}$ and the \gls{la} proportion of minority students $\overline{M}_{..k}$.  In model (i), on the left, the assumption is that school segregation patterns influence \gls{la} segregation patterns.  In model (ii), on the right, it is assumed that \gls{la} segregation patterns influence school segregation patterns. Note, however,  that there is no a-priori reason to make any of these assumptions,  as explained below.

The outcome model and the mediation equation for the school proportion of minority students $\overline{M}_{.jk}$   in  model (i) in Figure \ref{fg:medserial} are the same as in  \eqref{eq:contxla} and  \eqref{eq:modla} , respectively, for the parallel multiple mediation model.  The  mediation equation for the \gls{la} proportion of minority students $\overline{M}_{..k}$ then becomes

\begin{equation}\label{eq:memdser}
\overline{M}_{..k}  = \gamma_D + \omega^{D}_W M_{ijk} + \omega_S^{D}\overline{M}_{.jk}+ {e_{D}}_{..k} , \forall i,j \in k 
\end{equation}

However, it is not possible to estimate $ \omega_W^{D}$  from   \eqref{eq:memdser}.  Intuitively,  the district segregation $\omega^{D}$ is now estimated conditional on the school ethnic composition, that is, while holding constant the proportion of minority students $\overline{M}_{.jk}$  in the school each student $i$ attends,  effectively making $\omega_W^{D}$  the 'within-school district segregation'.  Since all students in the same school are also in the same \gls{la}, this implies that the \gls{la} proportion of minority students $\overline{M}_{..k}$ is also held constant when attempting to estimate $ \omega_W^{D}$.  With no variation in $\overline{M}_{..k}$,  $ \omega_W^{D}$ cannot be estimated.  In other words,  conditional on $\overline{M}_{.jk}$, the covariance between $M_{ijk}$ and $\overline{M}_{..k} $ is zero, because  $\overline{M}_{..k} $ is a constant, and hence  $\omega_W^{D} =0$.  

The outcome and mediation equation for the \gls{la} proportion of minority students $\overline{M}_{..k}$  in model (ii) in Figure \ref{fg:medserial} are the same as in  \eqref{eq:contxla} and  \eqref{eq:modla} ,  respectively, for the parallel multiple mediation model. The mediation equation for the school proportion of minority students $\overline{M}_{.jk}$ becomes

\begin{equation}\label{eq:memdserv}
\overline{M}_{.jk}  = \gamma_S + \omega^{S}_W M_{ijk} + \omega_D^{S}\overline{M}_{..k}+ {e_{S}}_{.jk} , \forall i,j \in k 
\end{equation}

In this case, it is possible to estimate the within-district school segregation parameter  $\omega^{S}_W$ and  the contextual effect of the \gls{la} proportion of minority students $\omega_D^{S}$.  Using the mediation analysis framework, this model allows one to decompose the overall ethnic achievement gap $\beta$ into one direct effect, $\beta^W$ and three indirect effects: $\omega^{S}_W\beta^{CS}$,  $\omega^D\beta^{CD}$ and $\omega_D^{S}\omega^D\beta^{CD}$.   In terms of the three decomposition approaches, this results in:

\begin{equation*}\label{eq:serialapp}
\begin{array}{rl}
  & \text{Approach 1} \\
\beta =& \beta^W + \left( \omega^{S}_W +\omega_D^{S}\omega^D\right) \beta^{CS} + \omega^{D}\beta^{CD} \\
 & \\
   & \text{Approach 2}  \\
\beta =& \left( 1-  \omega^{S}_W-\omega_D^{S}\omega^D\right) \beta^W +\left( \omega^{S}_W +\omega_D^{S}\omega^D-\omega^{D}\right)\beta^{BS} + \omega^{D}\beta^{BD} \\
 & \\
   & \text{Approach 3}  \\
\beta=& \left(1- \omega^{S}_W -\omega_D^{S}\omega^D\right) \beta^W + \left( \omega^{S}_W +\omega_D^{S}\omega^D-\omega^D\right)\beta^{W}+ \left( \omega^{S}_W +\omega_D^{S}\omega^D-\omega^D\right)\beta^{CS}  +\\
&  \omega^D\beta^W+ \omega^D\beta^{CS}+ \omega^D\beta^{CD}
\end{array}
\end{equation*}

Notice, however,  that $\omega^{S} = \omega^{S}_W +\omega_D^{S}\omega^D$, and hence all these decompositions are equivalent to those in section \ref{sx:multlevels},  derived using a parallel multiple mediation model.  Intuitively,  the serial multiple mediation model further decomposes the school segregation $\omega^{S}$ into a direct effect $ \omega^{S}_W$ and an indirect effect $\omega_D^{S}\omega^D$.  Arguably,  these equations for the decomposition approaches could be reparametrised to  further decompose the effects of school segregation and understand the contribution of the within-school segregation separately from the district contextual effect of school segregation and that would then be a motivation to use serial mediation models. 
 
\begin{figure}[htb]
\begin{center}
\includegraphics[width=0.8\textwidth]{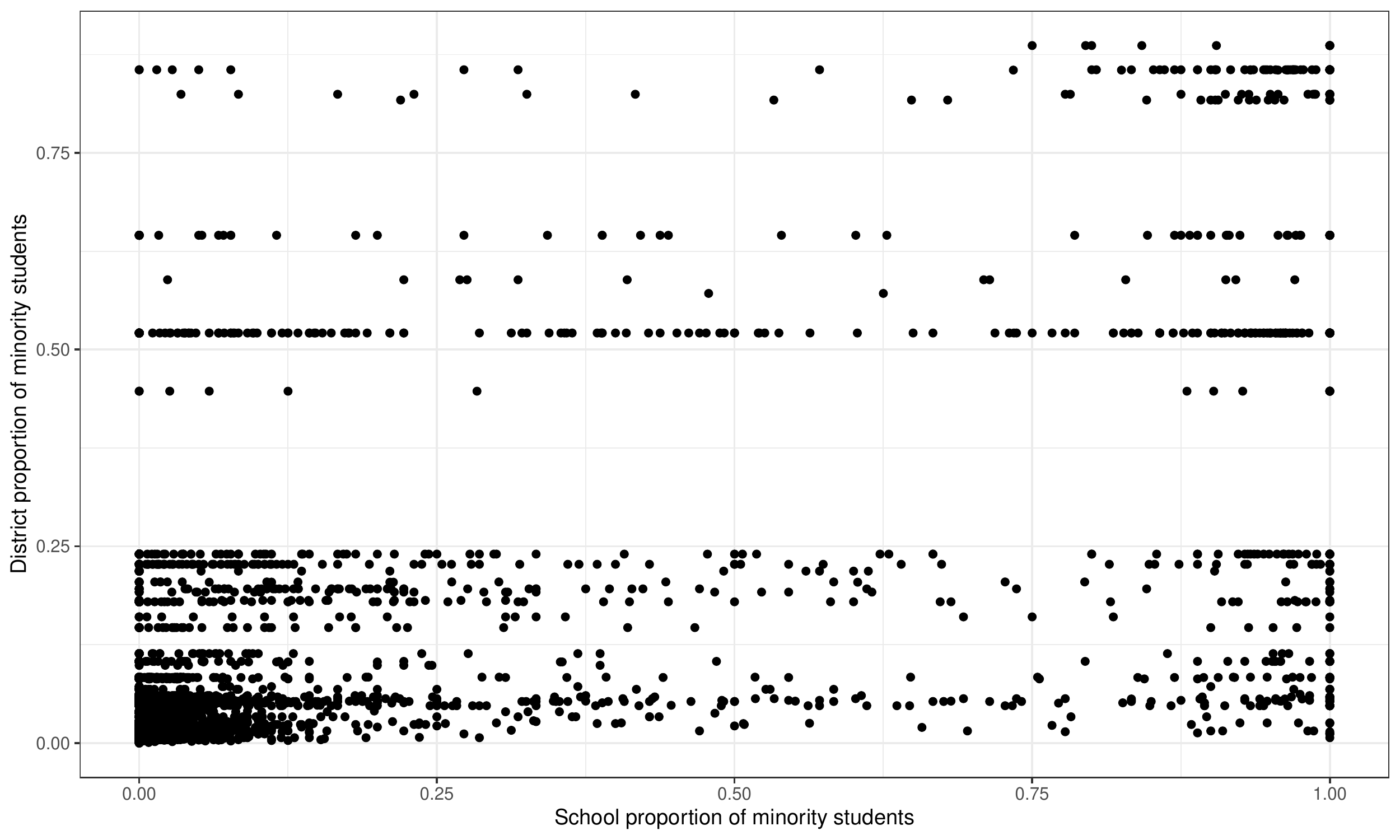} 
\caption{Relationship between the district  proportion of minority students and the school proportion of minority students.  Each point represents a school within a district}\label{fg:SnDMino}
\end{center}
\end{figure}

Notice, however, that the relation between the \gls{la} and school ethnic composition is only deterministic when all \glspl{la} have the same number of schools and the schools within these \glspl{la} have the same size.  Nonetheless, in the more conventional situation in which schools within \glspl{la} have different sizes, there is no deterministic relationship between these two;  \glspl{la} with the same proportion of minority students $0<\overline{M}_{..k}<1$ can have schools with proportion $0\leq\overline{M}_{.jk}\leq1$ in different shares.  Equivalently,  schools with the same proportion of minority students $0\leq\overline{M}_{.jk}\leq1$ can be located in \glspl{la} with very different proportions of minority students  $0<\overline{M}_{..k}<1$.  This is the case for Colombia, where,  as shown in Figure \ref{fg:SnDMino}, there is no clear relationship between the ethnic composition of schools and \glspl{la}. 

We encourage researchers to take advantage of the mediation framework to explore the potential extensions that the mediation framework offers, using either parallel multiple mediation models or  serial multiple mediation models as  best suits their research context and their research questions.  If,  in a particular context there is a reason to believe that \gls{la} segregation patterns influence school segregation patterns,  using serial multiple mediation models and reparametrising the decomposition approaches accordingly might result in improved insights to the ethnic achievement gaps in that particular context. Otherwise, extending the decomposition approaches using parallel multiple mediation models is likely to provide enough insights about the contribution of within-school and between-school differences to the overall ethnic achievement gap.  

\section{Discussion and Conclusion}\label{sx:agdisc}
In this paper we have examined three different approaches for a multilevel decomposition of the ethnic achievement gap. We argued that in Approach 1, \citet{Cook2000} and \citet{Fryer2006a,Fryer2004a} decomposed the Black-White achievement gap into the within-school gap  and the effect of segregation. In  Approach 2, \citet{Hanushek2007} decomposed the gap into a part that is attributable to the within-school gap and a part that is attributed to the between-school gap. Finally, in Approach 3, \citet{Reardon2008a} and  \citet{Page2008} decomposed the gap into three parts: one that is attributable to the within-school gap, a second one that is linked to the effect of school segregation through differences in student intake and a third component that is attributed to the effect of school segregation through differences in school composition.

Each of these decomposition approaches is useful to inform different research questions and policy decisions. For example, if the debate is about school segregation,  Approach 1 provides a more direct way to analyse its potential effects on the achievement gap. If the focus is on the within-school and between school components of the gap, Approach 2  is appropriate. If, in turn, the interest is in the mechanisms behind the between-school component of the gap, Approach 3  can be used.

The initial approaches were however limited in that they did not appropriately consider the role of additional levels of the school system (such as \glspl{la}) or multiple ethnic groups (beyond the binary White-minority comparison), which restricts the kind of policy recommendations that follow from the application of these methods.  We addressed these limitations by extending the three decomposition approaches to consider multiple levels of analysis and for multiple ethnic groups.

The role of between-\gls{la} differences (or the equivalent in the US, school districts) has only before been considered using \citet{Cook2000}'s method of including school- (in this case, \gls{la}) fixed effects into a Kitagawa-Oaxaca-Blinder decomposition.  \citet{Arteaga2017a} proposed using this method to examine the extent to which the gap between Indigenous and non-Indigenous students could be attributed to community effects in Peru. The use of the  Kitagawa-Oaxaca-Blinder \citep{Oaxaca1973,Blinder1973BW,Kitagawa1955} decomposition implies a different kind of decomposition than the one we explored here. 

Similarly, \citet{Dustmann2008} considered different ethnic groups by using Approach 1. In turn, \citet{Quinn2015a} used Approach 3 to decompose the Black-White gap in the US and included dummy variables for other ethnic groups to ensure they included all possible observations while comparing Black and White students (instead of Black and non-Black students). Nonetheless, they did not examine the role that other minority students play in explaining the Black-White achievement gap, which is also a component of the Black-White gap, as explained in section \ref{sx:multgroups}. Using Colombian data for illustration, we showed that the importance of this omission depends on the magnitude of the contextual effects and the relative exposure to other ethnic groups. In Colombia, despite strong contextual effects of Afrocolombian and Indigenous students, the little differential exposure to other minority groups imply that this is a small component of the gaps.

This illustration also showed the potential of the two extensions to transform the discussion around the ethnic achievement gaps. Considering \glspl{la} in the decomposition showed they play a role at least as important as schools. Therefore, it is worth incorporating them in the discussion about ethnic achievement gaps. Similarly, considering multiple ethnic groups showed that there is important heterogeneity among ethnic minority groups that is ignored when treating ethnic minorities as a single group. More importantly, the extension to the decomposition brings the discussion about segregation among minority groups to the forefront of the conversation about ethnic achievement gaps.

We showed that the existing decomposition approaches are mathematically equivalent to a mediation problem and used this equivalence to extent these approaches. Our extensions consider multiple levels and multiple ethnic groups, but this equivalence can be used to consider a wider range of extensions. For example, extending the decompositions to understand if they are different in private or public schools could be modelled as a moderated mediation model.  We also showed in section \ref{sx:serialmed} that serial multiple mediation models offer the possibility to further extent these decomposition approaches in contexts where there is a reason to believe that school segregation is influenced by \gls{la} segregation.

\subsection{Limitations}
The methodological discussion about the ethnic gap decomposition is tailored to its substantive application to the Colombian context, which implies that it only considers three levels that are observed in the data: students, schools and \glspl{la} and three ethnic minority groups. This restricts the detailed discussion of additional levels (such as cohorts and classrooms) and ethnic groups (which in England, for example, are usually reported using many more categories than in Colombia or the US). Nonetheless, the decomposition method can be generalised to these additional levels and groups. 

It is possible that the structure of the data is more complex than the hierarchical nesting of students within schools within \glspl{la} studied here.  For example, neighbourhoods may also be a relevant level to consider and students from multiple neighbourhoods may attend the same school but not all students from the same neighbourhood may attend the same school.  If a cross-classified (like in the example)  or  multiple membership  structure  provides a better representation of the ethnic achievement gaps,  the decomposition in this paper may overstate the contribution of schools and \glspl{la} to the overall ethnic achievement gap \citep{Goldstein1994CCM,BrowneGR2001MM,Leckie2013MMMLM,Martinez2012,Leckie2013CCMLM}.

Another limitation of the method for decomposing the ethnic achievement gap, as presented here, is that it does not allow for any interactions. Therefore, the models assume that the contextual effect of each minority group is the same for all ethnic groups or, equivalently, that the within-school gaps are the same regardless of the proportion of minority students in the schools and \glspl{la}. In the context of the Kitagawa-Oaxaca-Blinder decomposition, which \citet{Hou2014A} showed also fits under the mediation analysis framework, this is resolved by estimating separate equations for each group, which is an approach that can be further explored to expand the decomposition approaches in this paper. Additionally, potential equivalences with decomposition techniques that have been proposed for the analysis of segregation \citep[e.g.][]{Yamaguchi2017} merit further analysis but are out of the scope of this paper.

We showed that the current two-group two-level ethnic achievement gap decomposition approaches can be reformulated as a mediation problem. We used this equivalence to extent the existing decomposition approaches to consider multiple group and multiple level settings.  These extensions have the potential to provide important insights and hence policy implications for the reduction of these gaps, as illustrated by the application to Colombian data. Mediation analysis provides a flexible framework to work within. We encourage researchers to incorporate the achievement gap decomposition as a descriptive step in their analysis of ethnic achievement gaps and to take advantage of the unifying nature of the mediation analysis framework to tailor their own decomposition approaches.

 \pagestyle{plain}
 \theendnotes



\clearpage
\begin{appendices} 
\section{Generalisation of the Decomposition Approaches}\label{ax:ext}
\begin{figure}[htb]
\begin{center}
%
\includegraphics[width=0.25\textwidth]{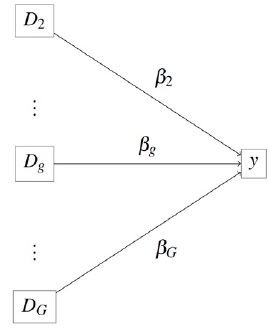}
\caption{Generalised total effect model for $G$ groups and $L$ levels}\label{fg:axagtotalgen}
\end{center}
\end{figure}

\begin{figure}[htb]
	\begin{center}
\includegraphics[width=0.75\textwidth]{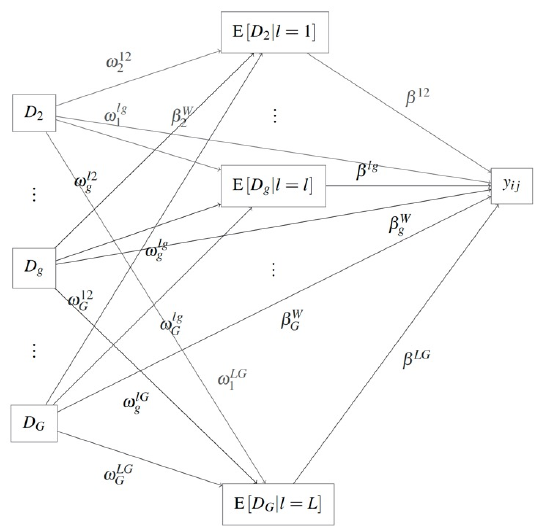}
	\caption{Generalised outcome and mediation models for $G$ groups and $L$ levels}\label{fg:axagoutmedgen}
\end{center}
\end{figure}

A generalised version of the model that allows decomposing the overall gap of $G$ groups into $L$ levels  is shown in figures \ref{fg:axagtotalgen} and \ref{fg:axagoutmedgen}.

In the general case, the categories of the independent variable (groups) are indexed by $g= 1,  \ldots ,G$ (where 1 is taken as the reference group), and the levels are indexed by $l=1,  \ldots ,L$. The $ijk\ldots$ subscripts have been omitted to avoid confusion. The variables $y$ and $D_g$ always vary by individual (level 1) and  the mediators vary at level $l$. The subscripts of the coefficients indicate the variable they multiply within each equation. The subscripts indicate effects that the group $g$ `receives' (e.g the overall gap $\beta_g$), while superscripts indicate effects that the group `produces' (e.g. the contextual effect of group $g$ at level $l$ $\beta^{lg}$).

The overall gaps are given by 
\begin{equation}\label{eq:gapxglaG}
y  = \alpha + \sum_{g=2}^{G}\beta_g {D_g}+{e_y}
\end{equation}

The outcome model is equivalent to the contextual effect model
\begin{equation}\label{eq:contxglaG}
y  = \alpha + \sum_{g=2}^{G}\beta_g^{W} {D_g}+ \sum_{l=1}^L{\sum_{g=2}^{G} \beta^{lg} \Ev{{D_g}| l=l}}+{e_y}
\end{equation}

Therefore, the $(G-1) \times L$ mediation equations are
\begin{equation}\label{eq:modglaG}
\begin{array}{rcl}
\Ev{D_2| l=1} & = &\omega_{012} + \sum_{g'=2}^G \omega_{g'}^{12} {D_{g'}}+ e_{12} \\
 & \vdots & \\
\Ev{D_{G}| l=1} & = &\omega_{01G} + \sum_{g'=2}^G {\omega_{g'}^{1G} D_{g'} }+ e_{1G} \\
 & \vdots & \\

\Ev{{D_g}| l=l} & = &\omega_{0lg} + \sum_{g'=2}^G \omega_{g'}^{lg} {D_{g'}}+ e_{lg} \\
& \vdots & \\
\Ev{{D_G}| l=L} & = &\omega_{0LG} + \sum_{g'=2}^G \omega_{g'}^{LG} {D_{g'}}+ e_{LG} \\

\end{array}
\end{equation}

\noindent and the $G-1$ gaps can be decomposed into multiple levels as

Approach 1:
\begin{equation}
\beta_g = \beta_g^W + \sum_{l=1}^{L}{\sum_g'{\omega_g^{lg'}\beta^{lg'}}},g'=2,...,G
\end{equation}

Approach 2:
\begin{equation}\label{eq:odecglaG}
\beta_g = \left(1- \omega_g^{1g}\right) \beta_g^W + \sum_{l=2}^{L} { \left(\omega_g^{(l-1)g}-\omega_g^{lg}\right)\beta^{B(l-1)g}  }  + \omega_g^{Lg}\beta^{BLg}+ \sum_{l=1}^L{\sum_{g' \neq g}\omega_g^{lg'}\beta^{lg'}}
\end{equation}

\noindent where $\beta^{Blg}= \beta_g^W + \sum_{l=1}^L{\beta^{lg}}$.

Approach 3:
\begin{equation}
\beta_g = \left(1- \omega_g^{1g}\right) \beta_g^W + \sum_{l=1}^{L}{\sum_g'{\omega_g^{lg'}\beta^{lg'}}} +\sum_{l=2}^{L}{\left(\omega_g^{\left(l-1\right)g}-\omega_g^{lg}\right)\beta^W_g}+\omega_g^{Lg}\beta^W_g
\end{equation}

\end{appendices}

\end{document}